\documentclass[11pt,a4paper]{article}

\usepackage{amsfonts}
\usepackage{multirow}
\usepackage{color}
\usepackage{cite}
\usepackage{graphicx}
\usepackage{float}
\usepackage{amsmath}
\usepackage{subcaption}
\usepackage{comment}

\usepackage[colorlinks=true,linkcolor=black,citecolor=black]{hyperref}

\title{\textbf{Scattering of wobbling vortices}}

\author{
	A. Alonso-Izquierdo$^{(a,b)}$, A. Gonz\'alez-Parra$^{(b)}$, A. Wereszczynski$^{(c,d)}$
	\\ {\normalsize {\it $^{(a)}$ Departamento de Matematica
			Aplicada, Universidad de Salamanca,}} 
    \\{\normalsize{\it Casas del Parque 2, 37008, Salamanca, Spain}}
	\\ {\normalsize {\it $^{(b)}$ IUFFyM, Universidad de Salamanca, Plaza de la Merced 1, 37008, Salamanca, Spain}}
    \\ {\normalsize {\it $^{(c)}$ Institute of Theoretical Physics, Jagiellonian University, Lojasiewicza 11, Krak\'{o}w, Poland }}
    \\ {\normalsize {\it $^{(d)}$ International Institute for Sustainability with Knotted Chiral Meta Matter (WPI-SKCM$^{\; 2}$)}} 
    \\ {\normalsize {\it Hiroshima University, 1-3-1 Kagamiyama, Higashi-Hiroshima, Hiroshima 739-8526, JAPAN }}
}

\date{}

\begin{document}

\maketitle

\begin{abstract}
We investigate the dynamical role of internal vibrational modes in the Abelian Higgs model, focusing on how Derrick-type excitations modify vortex dynamics and scattering processes. We study the scattering of excited vortices and show that the interplay between spectral flow and mode excitation generates effective forces and enables resonant energy transfer between translational and internal degrees of freedom. As a result, vortex dynamics become strongly non-adiabatic, exhibiting super-elastic collisions, oscillatory dependence of the final state on initial conditions, and the emergence of fractal structures in scattering diagrams. Our results demonstrate that internal vibrational modes play a fundamental role in vortex interactions, going beyond the standard moduli space approximation and revealing a rich phenomenology driven by mode dynamics.
\end{abstract}

\section{Introduction}\label{sec:intro}

Topological solitons constitute one of the most universal manifestations of nonlinearity in field theory. These localized, finite-energy configurations arise as nonperturbative solutions of nonlinear equations and are stabilized by the topological structure of the vacuum manifold. Their presence spans a wide range of physical systems, from condensed matter to cosmology, making them a central object of study in modern mathematical physics \cite{manton2004, vilenkin1994}. Among the different classes of topological defects, vortices in the Abelian Higgs model represent the prototypical example in two spatial dimensions, providing a paradigmatic framework to investigate the interplay between gauge fields, symmetry breaking, and nonlinear dynamics \cite{weinberg2012}. In Type-II superconductors they describe quantized tubes of magnetic flux whose collective behavior leads to the formation of vortex lattices and rich phase structures. In quantum fluids and weakly interacting Bose gasses, vortices appear as macroscopic quantum states carrying quantized circulation, exhibiting a core structure and velocity fields closely analogous to those of classical hydrodynamics \cite{gross1961, pitaevskii1961, fetter1965}. In relativistic settings, vortex solutions are interpreted as planar cross-sections of cosmic strings which are expected to be formed during symmetry-breaking phase transitions in the early universe \cite{Hindmarsh}, where they may play a role in generating observable cosmological signatures \cite{vilenkin1994,blanco2025}. These diverse realizations highlight the universality of vortex phenomena and motivate their study across different theoretical frameworks.

The theoretical understanding of vortices in relativistic field theory originates from the seminal works of Abrikosov, Nielsen, and Olesen, which established the existence of finite-energy flux-tube solutions in Ginzburg–Landau and gauge field theories \cite{abrikosov1957, nielsen1973}. A key development was the identification of the Bogomolny–Prasad–Sommerfield (BPS) limit, in which the field equations reduce to a set of first-order self-dual equations \cite{bogomolny1976, prasad1975}. At this critical point, the scalar and vector boson masses coincide, and the static forces between vortices vanish, leading to a degenerate continuous set of solutions. In particular, configurations with an arbitrary winding number can be constructed as superpositions of elementary vortices located at arbitrary positions in the plane \cite{weinberg1979, taubes1980, jaffe81051}. This degeneracy gives rise to a natural geometric description of vortex dynamics in terms of the moduli space of static solutions, determined by the coordinates of the vortex centers. In the low-energy regime, the motion of vortices can be approximated by geodesic flow on this manifold, whose metric was explicitly constructed by Samols \cite{samols1992}. This approach provides a remarkably accurate description of slow vortex dynamics and captures characteristic phenomena such as the well-known $90^\circ$-scattering in head-on collisions of two vortices. This approximation has been found to reproduce the results of numerical simulations with remarkable accuracy \cite{shellard1988,ruback1988, rebbi1992}. Despite its success, however, the moduli space approximation is intrinsically limited to configurations that remain close to static solutions and, therefore, does not capture the effects of internal excitations. A more complete description of vortex dynamics requires the analysis of fluctuations around static configurations. The spectrum of small perturbations reveals the existence of internal or shape modes, which correspond to localized vibrational degrees of freedom of the vortex \cite{arodz1991, goodband1995, alonso2016a, alonso2016b, alonso2024a, alonso2026a}. A similar result is found for $\mathbb{CP}^1$-vortices, as shown in \cite{gavrea2026}. Among these, the Derrick-type (breathing) modes play a particularly important role as they describe oscillations in the size of the vortex core. The structure of the fluctuation spectrum becomes increasingly rich for multi-vortex configurations, where additional modes arise and their frequencies depend on the relative positions of the vortices. This dependence leads to the phenomenon of spectral flow, whereby the internal mode frequencies evolve dynamically as the vortices move in the configuration space \cite{alonso2024b}.

In recent years, it has become increasingly clear that the excitation of internal vibrational modes plays a crucial role in determining vortex dynamics. One of the earliest studies addressing this issue is \cite{krusch2024}, where it was shown that the scattering of excited vortices can lead to chaotic dynamics. Subsequent developments have provided a deeper understanding of the mechanisms underlying these phenomena. In particular, the collective coordinate approach developed in \cite{alonso2024c, miguelez2025} demonstrated that the spectral flow of the internal modes generates effective forces between vortices. These forces may either enhance or oppose the motion predicted by the geodesic approximation, depending on the relative phase of the excited shape modes. As a consequence, the dynamics deviates significantly from the force-free picture characteristic of the BPS regime. A particularly striking manifestation of this scenario is the emergence of spectral walls, as discussed in \cite{alonso2024d}. These structures act as dynamical barriers that can reflect or trap vortices during their evolution. Similar phenomena had previously been identified in the context of kink dynamics in $(1+1)$-dimensional scalar field theories \cite{adam2019}, highlighting a deep connection between soliton dynamics across different dimensions. These ideas have been further extended to multiple vortex configurations. The scattering of excited BPS 3-vortex systems has been investigated in \cite{alonso2025a}, while more general $n$-vortex configurations have been analyzed in \cite{bachmaier2026b}, where it was shown that mode-driven dynamics substantially deform the geodesic trajectories on moduli space.

Beyond the effective forces induced by internal modes, vortex collisions also involve nontrivial exchanges of energy between translational and vibrational degrees of freedom through resonant mechanisms. This interplay gives rise to strongly non-adiabatic dynamics, characterized by phenomena such as multi-bounce scattering and the emergence of fractal structures in the dependence of the final state on the initial conditions. Although these effects have been extensively studied and are well understood in lower-dimensional soliton systems \cite{campbell1983, sugiyama1979, alonso2021a, alonso2022a, manton2021, Azadeh2021, Blaschke2024}, a systematic and comprehensive investigation of this phenomenon in vortex dynamics is still lacking. Partial results have been obtained in \cite{Alonso2026c}, while its implications for vortex–antivortex collisions have only recently been addressed  \cite{bachmaier2026}. In this work, we undertake a detailed investigation of the dynamics of excited vortices in the Abelian Higgs model, focusing on configurations where Derrick-type internal modes are activated. Our aim is to elucidate how initial vibrational excitations modify vortex–vortex interactions and to uncover the underlying mechanisms governing the transfer of energy between different dynamical channels. By combining analytical insights with numerical simulations, we demonstrate that excited vortices exhibit a rich phenomenology that departs markedly from the standard geodesic description. In particular, we identify dynamical regimes in which resonant energy exchange leads to pronounced non-adiabatic effects, including super-elastic scattering events—where the outgoing vortices acquire greater kinetic energy than initially—and the formation of intricate fractal patterns in scattering observables. These results highlight the crucial role of internal degrees of freedom in shaping vortex dynamics and establish a new framework for understanding complex interactions in topological soliton systems.

The paper is organized as follows. In Sections 2 and 3 we introduce the theoretical framework and establish the notation and conventions used throughout the article. In particular, Section 2 is devoted to a brief review of the Abelian Higgs model and the main properties of vortex solutions, while Section 3 presents a detailed analysis of their internal vibrational spectrum, with special emphasis on the Derrick-type excitation of the single vortex. In Section 4 we investigate the scattering of excited vortices, focusing on how internal modes modify the interaction dynamics and on the role played by resonant energy transfer mechanisms. We will demonstrate that the Type-I, Type-II, and self-dual regimes leave distinct dynamical signatures on vortex scattering. These features are manifested in the structure of the scattering diagrams, which display both the final velocities of the outgoing vortices and their associated excitation amplitudes as functions of the initial collision velocity. Finally, Section 5 summarizes our main results and outlines possible directions for future research.

\section{The Abelian-Higgs model}

The Abelian Higgs model provides one of the most extensively studied frameworks for the description of topological defects in gauge field theories. It consists of a complex scalar field $\phi:\mathbb{R}^{1,2} \rightarrow \mathbb{C}$, $\phi(x)=\phi_1(x)+i\phi_2(x)$, minimally coupled to a $U(1)$ gauge field $A_\mu=(A_0(x),A_1(x),A_2(x))$, defined on a $(1+2)$-dimensional Minkowski spacetime. Using dimensionless variables, the dynamics of the system are derived from the action functional
\begin{equation}
	S[\phi,A]=\int d^3 x \left[ -\frac{1}{4} F_{\mu\nu}F^{\mu \nu} + \frac{1}{2} \overline{D_\mu \phi}\, D^\mu \phi -U(\phi,\overline{\phi}) \right]
	\label{action1} ,
\end{equation}
where the self-interaction potential term $U(\phi,\overline{\phi})$ is given by
\begin{equation}
	U(\phi,\overline{\phi}) = \frac{\lambda}{8} (\overline{\phi}\, \phi-1)^2  \label{potentialU} ,
\end{equation}
being $\lambda$ the non-trivial coupling constant of the system. The interaction between the scalar and gauge sectors is encoded through the covariant derivative $D_\mu \phi(x) = (\partial_\mu -i A_\mu(x))\phi(x)$ while the dynamics of the gauge field is governed by the electromagnetic field strength tensor $F_{\mu\nu}(x)=\partial_\mu A_\nu(x) - \partial_\nu A_\mu(x)$. Here and in the following, $\overline{\phi}$ denotes the complex conjugate of the scalar field. Throughout this work, we adopt the convention $\eta_{\mu\nu}={\rm diag}(1,-1,-1)$ for the Minkowski metric with $\mu,\nu=0,1,2$. We also employ the Einstein summation convention over repeated indices. Greek indices refer to spacetime components, whereas Latin indices are reserved for spatial coordinates. 

In the temporal gauge $A_0=0$, the field equations take the form of the coupled system of second-order nonlinear partial differential equations 
\begin{eqnarray}
	&& \frac{1}{2}\partial_0^2 \phi - \frac{1}{2}D_jD_j\phi = - \frac{\partial U }{\partial \overline{\phi}} \hspace{0.0cm}, \label{genedo1} \\
	&& \partial_0^2 A_j  - \partial_k F_{kj} = - \frac{i}{2} \Big[ \overline{\phi} \, D_j \phi -  \overline{D_j \phi} \,\phi \Big] \hspace{0.2cm}, \label{genedo2}
\end{eqnarray}
which must be complemented by the Gauss law constraint, 
\begin{equation}
	\partial_{0i}A_i = - \frac{i}{2} \Big[ \overline{\phi} \, \partial_0 \phi -  \overline{\partial_0 \phi} \,\phi \Big] \hspace{0.2cm}, \label{GaussLaw}
\end{equation}
which ensures the consistency of the gauge choice. For static configurations, this constraint is automatically satisfied, and the system reduces to a set of two partial differential equations that admit topologically nontrivial finite-energy smooth solutions known as vortices. From these requirements, the scalar field asymptotically approaches the vacuum configuration,
\begin{equation}
	\lim_{r\rightarrow \infty}\phi\Big|_{S^r~} = e^{i n \theta} \label{asymptotic1} \hspace{0.2cm},
\end{equation}
whereas the gauge field asymptotically becomes a pure gauge configuration
\begin{equation}
	\lim_{r\rightarrow \infty} (A_1,A_2)\Big|_{S^r} = \left(-i e^{-in\theta} \partial_1 e^{in\theta},-ie^{-in\theta} \partial_2 e^{in\theta} \right) \label{asymptotic2}
\end{equation}
with $n\in \mathbb{Z}$. In (\ref{asymptotic1}) and (\ref{asymptotic2}) polar coordinates, $(r,\theta)$, has been used on the spatial plane, defined as usual by $x_1=r\cos\theta$ and $x_2=r \sin \theta$. Thus, $S^r$ stands for a circular boundary of radius $r$.  As a consequence of this asymptotic behavior, the magnetic flux carried by the vortices is quantized 
\[
\Phi = \frac{1}{2\pi}  \int_{\mathbb{R}^2} d^2x\, F_{12} = n \in \mathbb{Z} 
\]
and the total vortex potential energy 
\begin{equation}
	V[\phi,A] = \int_{\mathbb{R}^2} d^2x \, \mathcal{E}[\phi,A]  \label{PotentialEnergy}
\end{equation}
remains finite. Indeed, the energy density of these  solutions (the integrand in (\ref{PotentialEnergy})),
\begin{equation}
	\mathcal{E}[\phi,A] = \frac{1}{2} F_{12}^2 + \frac{1}{2} \overline{D_i \phi} \, D_i \phi + \frac{\lambda}{8} (\overline{\phi} \phi -1)^2 \label{EnergyDensity}
\end{equation}
is localized around a point in the plane that is considered the vortex center.

Static rotationally invariant $n$-vortex solutions can be constructed by imposing the radial gauge condition $A_r=0$ and adopting the ansatz
\begin{equation}
	\phi(r,\theta) = f_n(r) e^{in\theta} \hspace{0.5cm},\hspace{0.5cm} A_\theta(r,\theta) = n \beta_n(r) \, , \label{ansatz}
\end{equation}
where we use the convention $A_1=A_r \cos \theta - A_\theta \sin \theta$ and $A_2=A_r \sin \theta + A_\theta \cos \theta$. Substituting the ansatz (\ref{ansatz}) into the field equations (\ref{genedo1}) and (\ref{genedo2}) reduces the problem to a system of ordinary differential equations for the functions $f_n(r)$ and $\beta_n(r)$, namely
\begin{align*}
	& \frac{d^2 f_n}{dr^2} + \frac{1}{r} \frac{d f_n}{dr} - \frac{n^2(1-\beta_n)^2 f_n}{r^2} + \frac{\lambda}{2} f_n  (1-f_n^2) = 0 \\
	& \frac{d^2 \beta_n}{dr^2} - \frac{1}{r} \frac{d \beta_n}{dr}+(1-\beta_n ) f_n^2 = 0 \label{bpsedo}
\end{align*}
In addition, regularity at the origin imposes boundary conditions on the profile functions, which must satisfy
\[
f_n(0) = 0 \hspace{0.5cm}\mbox{and}\hspace{0.5cm} \beta_n(0)=0  .
\]
By analyzing the previous equations near $r=0$, it can be shown that the profile functions $f_r$ and $\beta_n$ admit a power series expansion around the origin of the form
\[
	f_n(r) = r^n \Big( d_0^{(n)} - d_0^{(n)} \frac{2n^2 c_0^{(n)} + \frac{\lambda}{2}}{4(n+1)} r^2 + \dots \Big) \hspace{0.0cm},\hspace{0.5cm} \beta_n(r) = r^2 \Big( c_0^{(n)} - \frac{(d_0^{(n)})^2}{4n(n+1)} r^{2n} + \dots \Big)  \hspace{0.2cm}, 
\]
where $d_0^{(n)}$ and $c_0^{(n)}$ are real constants that depend on the topological charge $n$ of the vortex configuration and on the coupling constant $\lambda$. Therefore, the dominant terms for the 1-vortex (with $n=1$) at $r\approx 0$ are given by
\begin{equation}
f_{n=1}(r) \approx \overline{d}_0 \, r \hspace{0.5cm},\hspace{0.5cm} \beta_{n=1}(r) \approx \overline{c}_0 \, r^2\hspace{0.5cm},\label{Seriefyb}
\end{equation}
where we have denoted $\overline{d}_0=d_0^{(1)}$ and $\overline{c}_0=c_0^{(1)}$ for simplicity.

On the other hand, the asymptotic conditions for these solutions are
\begin{equation}
	\lim_{r\rightarrow \infty} f_n(r)= 1 \hspace{0.0cm},\hspace{0.5cm} \lim_{r\rightarrow \infty} \beta_n(r) = 1 \, .  \label{asymptotic}
\end{equation} 
In order to obtain explicitly the profiles $f_n(r)$ and $\beta_n(r)$ for every value of $r$, equations (\ref{genedo1}) and (\ref{genedo2}) must be integrated numerically. It is worth emphasizing that the energy density $\mathcal{E}_n(r)=\mathcal{E}[f_n(r),\beta_n(r)]$ of a 1-vortex attains its maximum at the vortex center. In contrast, for higher winding numbers ($n>1$), the energy density no longer peaks at the origin but instead reaches its maximum along a circular ring located at a finite distance from the vortex center.

\section{Derrick-type vibrational modes of vortices}

\label{bps}

As outlined above, our primary goal is to investigate the scattering of two 1-vortices excited through the Derrick-type mode, that is, two wobbling $1$-vortices. To address this problem, we briefly review the spectral structure of $n$-vortex configurations in the Abelian Higgs model for arbitrary values of the coupling constant $\lambda$. A comprehensive analysis of this spectral problem can be found in \cite{alonso2026a}. To facilitate the analysis, we introduce a specific notation for the static, rotationally symmetric $n$-vortex solutions. In particular, we denote the profiles of the scalar and gauge fields by
\[
\psi(x_i;n)=\psi_1(x_i;n) + i \, \psi_2(x_i;n) \hspace{00cm},\hspace{0.5cm} V(x_i;n)=(V_1(x_i;n),V_2(x_i;n)) \hspace{0.5cm}\mbox{with}\hspace{0.5cm} i=1,2 ,
\]
which will be assembled into a four real-component column $\Psi(r,\theta) \in {\cal C}^\infty(\mathbb{R}^2) \oplus \mathbb{R}^4$ in the form
\begin{equation}
	\Psi(r,\theta)=\left(\begin{array}{c} V_1(x_i;n) \\ V_2(x_i;n) \\ \psi_1(x_i;n) \\ \psi_2(x_i;n) \end{array}  \right) =  \left(\begin{array}{c}- \frac{n\beta_n(r)}{r} \sin \theta \\ \frac{n\beta_n(r)}{r} \cos \theta  \\ f_n(r) \cos (n\theta) \\ f_n(r) \sin (n\theta) \end{array}  \right) \label{vortexcolumn}.
\end{equation}
Consequently, a perturbed $n$-vortex  $\widetilde{\Psi}(x_i,n)$ can be expressed in the form
\begin{equation}
	\widetilde{\Psi}(x_i,n) = \Psi(x_i,n) + \eta \, \xi(x_i)=\left(\begin{array}{c} V_1(x_i;n) \\ V_2(x_i;n) \\ \psi_1(x_i;n) \\ \psi_2(x_i;n)\end{array}  \right) + \eta \,  \left(\begin{array}{c} a_1(x_i) \\ a_2(x_i)  \\ \varphi_1(x_i) \\ \varphi_2(x_i) \end{array}  \right) \label{perturvortexcolumn},
\end{equation}
where 
\[
\xi(x_i)=\left( \begin{array}{c c c c}a_1(x_i) & a_2(x_i) & \varphi_1(x_i) & \varphi_2(x_i) \end{array} \right)^t ,
\]
denote the fluctuation column and $\eta$ is the amplitude of the perturbation. As usual in this context the \textit{background gauge}
\begin{equation}
	\partial_1 a_1( x_i) + \partial_2 a_2( x_i) - (\,\psi_1( x_i)\, \varphi_2( x_i)-\psi_2( x_i)\,\varphi_1( x_i)\,)=0  
	\label{backgroundgauge}
\end{equation}
is imposed as the gauge fixing condition on the fluctuation modes. With this set-up the normal modes of an $n$-vortex solution $\Psi(x_i,n)$ are determined by the spectral condition 
\begin{equation}
	{\cal H}^+ \xi_\nu(x_i) =\omega_\nu^2 \, \xi_\nu(x_i) , \label{spectralproblem}
\end{equation}
where $\nu$ is a label in either the discrete or the continuous spectrum useful to enumerate the eigenfunctions and eigenvalues, and ${\cal H}^+$ is the second-order small fluctuation operator
{\small\begin{equation}
		{\cal H}^+= \left( \begin{array}{cccc}
			-\Delta + |\psi|^2 & 0 & -2 \widetilde{D}_1 \psi_2 & 2 \widetilde{D}_1 \psi_1 \\
			0 & -\Delta +|\psi|^2 & -2 \widetilde{D}_2 \psi_2 & 2 \widetilde{D}_2 \psi_1 \\
			-2 \widetilde{D}_1 \psi_2 & -2 \widetilde{D}_2\psi_2 & -\Delta + \frac{3\lambda}{2} \psi_1^2 + (1+\frac{\lambda}{2})\psi_2^2 -\frac{\lambda}{2}  +V_kV_k & -2 V_k \partial_k + (\lambda-1)\psi_1\psi_2\\
			2\widetilde{D}_1\psi_1 & 2 \widetilde{D}_2 \psi_1 & 2V_k \partial_k + (\lambda-1)\psi_1\psi_2 & -\Delta + (1+\frac{\lambda}{2})\psi_1^2 +\frac{3\lambda}{2} \psi_2^2  -\frac{\lambda}{2} + V_kV_k
		\end{array} \right)\label{hessianoperator}
\end{equation}}
where $\widetilde{D}_i\psi_j = \partial_i\psi_j+\epsilon^{ik} V_i \psi_k$. The Sturm-Liouville eigenvalue problem (\ref{spectralproblem}) is obtained by linearizing the field equations (in the background gauge) around the vortices, see \cite{alonso2016a}. 

In particular, we investigate the Derrick-type vibrational modes associated with vortex solutions. These modes are distinguished by the fact that they preserve the angular dependence of the underlying static configuration. Therefore, the Derrick-type modes follow the form 
\begin{equation}
	\xi_\nu(x_i,n) = \left( \begin{array}{c} v_n(r) \, \sin \theta  \\ - v_n(r) \,\cos \theta  \\ u_n(r) \, \cos(n\theta) \\ u_n(r) \, \sin(n\theta) 
	\end{array} \right),  \label{genericform03b}
\end{equation}
where the radial functions $v_n(r)$ and $u_n(r)$ depend on the vorticity of the solution. Upon substitution into the spectral problem \eqref{spectralproblem}, this ansatz reduces the full system to an effective eigenvalue problem for the radial profiles $v_n(r)$ and $u_n(r)$. Consequently, the original problem simplifies to
\begin{equation}
	{\cal H}^+_R \zeta_\nu (x) = \omega_\nu^2 \zeta_\nu(x) \, , \label{reducedSpectral}
\end{equation}
where the radial functions comprise the column $\zeta_\nu (r)= (v_n(r),u_n(r) )^t$ and the operator ${\cal H}^+_R$ is given by
\begin{equation}
	{\cal H}^+_R  = \left( \begin{array}{cc} - \frac{d^2}{dr^2} - \frac{1}{r} \frac{d}{dr} + \frac{1}{r^2} + f_n^2(r)  &  \frac{2n}{r} (1-\beta_n(r)) f_n(r) \\ \frac{2n}{r} (1-\beta_n(r)) f_n(r) & - \frac{d^2}{dr^2}- \frac{1}{r} \frac{d}{dr} + \frac{n^2 (1-\beta_n(r))^2}{r^2} + \frac{3\lambda}{2} f_n^2(r) - \frac{\lambda}{2} 
	\end{array} \right). \label{operatorhbar0}
\end{equation}

In this work, we focus on the scattering of excited 1-vortex configurations and, for this reason, we analyze the spectrum of small fluctuations obtained from the spectral problem \eqref{operatorhbar0} for $n=1$. In this case, the single 1-vortex supports a unique localized bound mode when $\lambda <1.5$. This mode corresponds to a radial (breathing) excitation, which induces periodic variations in the size of the vortex core. In linear order, the associated perturbation can be described by an eigenfunction of the form
\[
\xi(x_i,1) \, e^{i\omega t} = \left( \begin{array}{c} v_{n=1}(r) \\ u_{n=1}(r) \end{array} \right) e^{i\omega t} 
\] 
where $\omega$ denotes the corresponding eigenfrequency and, as usual, $t\equiv x_0$. In the next section, we will also employ the notation $x\equiv x_1$ and $y\equiv x_2$ for simplicity. The resulting motion is therefore periodic, with a characteristic timescale given by  $T=\frac{2\pi}{\omega}$. Physically, this mode manifests as a coherent oscillation in which the vortex core undergoes successive expansions and contractions. As a consequence, the energy density at the center of the vortex does not remain constant but instead oscillates in time, reflecting the underlying vibrational dynamics of the configuration.

Since the energy density of the single vortex attains its maximum at the vortex center, analysis of its oscillatory behavior can be effectively performed by examining the behavior of the vibrational modes in the core region. Using the Frobenius method applied to the differential equations derived from (\ref{reducedSpectral}) leads to expressions
\begin{equation}
	v_{n=1}(r) \approx \overline{v}_0 \, r \hspace{0.5cm},\hspace{0.5cm} u_{n=1}(r) \approx \overline{u}_0 \, r \label{DerrickApproximation}
\end{equation}
for the radial profiles in the limit $r\approx 0$ where the values of $\overline{v}_0$ and $\overline{u}_0$ depend on the coupling constant $\lambda$ and must be derived numerically. The expressions in \eqref{DerrickApproximation} provide a convenient framework to analyze the time evolution of the energy-density peak of the 1-vortex configuration under a Derrick-type excitation. Substituting \eqref{Seriefyb} and \eqref{DerrickApproximation} into the linearly perturbed vortex configuration \eqref{perturvortexcolumn}, and subsequently into the energy density \eqref{EnergyDensity}, allows us to extract the leading contribution at the vortex center. Taking the limit $r\approx 0$, we obtain an explicit expression for the time dependence of the maximum of the energy density, which takes the form
\begin{equation}
{\cal E}_{\rm max}(t) = \frac{1}{8} \Big( \lambda +16 \overline{c}_0^2+ 8 \overline{d}_0^2 \Big) + \frac{1}{2} \Big( 4 \,\overline{d}_0\, \overline{u}_0- 8\, \overline{c}_0\, \overline{v}_0 \Big)\eta \sin (\omega t) + \frac{1}{4} \Big( 4 \overline{u}_0^2 + 8 \overline{v}_0^2 \Big) \eta^2 \sin^2 (\omega t)\label{EnergyDensityPeak}
\end{equation}
In this expression, $\eta$ denotes the amplitude of the perturbation appearing in  (\ref{perturvortexcolumn}), assuming that the global eigenfunction $\xi(x_i)$ is normalized over the plane, that is, $\|\xi(x_i)\|=\int_{\mathbb{R}^2} d^2x [ (a_1(x_i))^2+ (a_2(x_i))^2 + (\varphi_1(x_i))^2 + (\varphi_2(x_i))^2] =1$. From expression (\ref{EnergyDensityPeak}) it follows that the average value of the peak energy density is given by the amplitude-dependent formula 
\[
\overline{\cal E}_{\rm max} = \frac{\lambda}{8} + 2 \overline{c}_0^2 + \overline{d}_0^2 + \frac{1}{2} \eta^2 (\overline{u}_{0}^2 + 2 \overline{v}_{0}^2 )
\]
while the critical points of this time-dependent function ${\cal E}_{\rm max}(t) $ occur at 
\[
t_1(k)  = \pm \frac{\pi}{2\omega} + \frac{2\pi}{\omega} k \hspace{0.5cm}, \hspace{0.5cm} k\in \mathbb{Z} \hspace{0.2cm},
\]
for moderate values of the excitation. The oscillatory behavior of the peak energy density can be characterized by considering the difference between the maximum and minimum values reached during one oscillation cycle. Using the expressions for the maxima and the minima occurring at the times $t_1(k)$, one obtains
\begin{equation}
	\Delta {\cal E}_{\rm max} = 4 \eta \, | 2 \overline{c}_0 \overline{v}_{0}-\overline{d}_0 \overline{u}_{0} |  .\label{DeltaE}
\end{equation}
The expressions derived previously are valid under the assumption that the vortex center remains fixed in space. However, in this work we are interested in the collision of excited 1-vortices, where motion plays a fundamental role. Since the Abelian Higgs model defined in (\ref{action1}) is Lorentz invariant, the static vortex solutions constructed in the previous sections can be converted to traveling configurations by means of a Lorentz boost. When a vortex moves with velocity $v$, relativistic effects must be taken into account. By extending the analysis performed in the static case to a boosted configuration, the expression (\ref{EnergyDensityPeak}) becomes
\begin{align}
{\cal E}_{\rm max}(t) &= \frac{1}{8(1-v^2)} \Big[16 \overline{c}_0^2+8 \overline{d}_0^2 + \lambda + v^2 \Big( 4 \overline{c}_0^2 - 4 \overline{d}_0^2 - \lambda \Big) - \nonumber \\
& - 8 \Big( -2(\overline{d}_0 \overline{u}_{0} - 2 \overline{c}_0 \overline{v}_{0}) + v^2 (\overline{d}_0 \overline{u}_{0}+ \overline{c}_0 \overline{v}_{0}) \Big) \eta \sin (t \omega \sqrt{1-v^2}) - \label{GeneralAmplitude} \\
& - 4 \Big( -2 \overline{u}_{0}^2 - 4 \overline{v}_{0}^2 + v^2 (\overline{u}_{0}^2- \overline{v}_{0}^2) \Big) \eta^2 \sin^2 (t\omega \sqrt{1-v^2}) \Big] \hspace{0.2cm}. \nonumber
\end{align}
Additionally, the transverse Doppler effect modifies the oscillations of the energy-density peak described by (\ref{DeltaE}) in the form 
\begin{equation}
	\Delta {\cal E}_{\rm max}(v) = \frac{2\eta}{1-v^2}\Big| 2(2\overline{c}_0 \overline{v}_{0}-\overline{d}_0 \overline{u}_{0}) + v^2(\overline{d}_0 \overline{u}_{0}+\overline{c}_0 \overline{v}_{0}) \Big| \, , \label{DeltaE2}
\end{equation}
which depends on the velocity $v$ of the traveling vortex. This relation offers a straightforward method to determine the excitation amplitude from the variation of the peak energy density:
\begin{equation}
\eta 	= \frac{(1-v^2)\Delta {\cal E}_{\rm max}}{|4(2\overline{c}_0 \overline{v}_{0}-\overline{d}_0 \overline{u}_{0}) + 2v^2(\overline{d}_0 \overline{u}_{0}+\overline{c}_0 \overline{v}_{0})| } \, . \label{AmplitudE2}
\end{equation}

\section{Scattering of excited vortices}

In this section, we analyze the scattering of two 1-vortices undergoing a head-on collision when both are initially excited in phase. The initial condition is determined by a symmetric configuration consisting of two vortices separated by a distance $2d$ along the $x$-axis. Each vortex is excited through its Derrick-type mode with amplitude $\eta_0$, and they are pushed toward each other with an initial velocity $(\pm v_0,0)$. Due to the symmetry of the initial configuration under the transformations $x\rightarrow -x$ and $y\rightarrow -y$, one expects that, for sufficiently small excitation amplitudes or sufficiently large initial velocities, the vortices collide at the origin of the plane. After the interaction, they typically undergo the characteristic $90^\circ$-scattering, moving apart from each other along the $y$-axis with a final velocity $v_f=(0,v_y)$ and a modified excitation amplitude $\eta_f$. However, this behavior can be significantly altered depending on the model regime, the initial velocity and the excitation amplitude. In particular, away from the self-dual regime, static intersolitonic forces play an important role \cite{Speight1997}. For $\lambda<1$, vortices experience attractive forces, which may lead to the formation of quasi-bound states characterized by repeated $90^\circ$-scattering events. Conversely, for $\lambda>1$, vortices repel each other, and for sufficiently low collision velocities the interaction may prevent the vortices from reaching the collision point, resulting instead in their separation along the original direction of motion. In addition to these forces, the presence of in-phase Derrick-type excitations introduces a new effective attractive interaction arising from the spectral flow of internal modes during the evolution \cite{alonso2024c}. This mechanism further modifies the scattering dynamics, leading to a nontrivial interplay between vibrational and translational degrees of freedom.

The main goal of this section is to provide a global picture of how these different mechanisms affect the scattering of two 1-vortices. To this end, we perform a systematic set of numerical simulations and present the results in a graphical form that allows for a clear identification of the dominant dynamical features. In particular, we analyze the dependence of the final vortex velocity $v_f$ (measured once the vortices are sufficiently separated and no longer interact with each other) on the initial collision velocity $v_0$. This defines the so-called \textit{velocity diagram}. In addition, we study the dependence of the final excitation amplitude on the initial velocity, providing us \textit{the amplitude diagram}. In practice, we present separate diagrams for the two components of the final velocity, namely $v_f=(v_x,v_y)$. It is important to note that the direction of separation depends on the number of effective $90^\circ$-scattering events. If an odd number of such events occurs, the vortices separate along the $y$-axis, implying $v_x=0$. Conversely, if the number is even, the vortices separate along the $x$-axis, and therefore $v_y=0$. It should also be emphasized that certain scattering processes may lead to the formation of long-lived quasi-bound states, in which the vortices undergo repeated bounces without clearly separating within the timescale of the simulation. In such cases, we assign a vanishing final velocity, $v_f=(0,0)$. An analogous prescription is adopted for the final excitation amplitude. When the vortices do not separate asymptotically, we set it to zero. With this convention, such events appear as gaps in the graphical representations shown throughout the paper.

These scattering diagrams provide valuable insight into the role of resonant energy transfer mechanisms in vortex scattering, a phenomenon that has been extensively studied in kink–antikink collisions in the $\phi^4$-model \cite{alonso2021a, alonso2022a}. In contrast to this case, where attractive forces are always present, the Abelian Higgs model offers a natural framework in which the influence of intersolitonic forces can be systematically explored. Indeed, it allows for the comparison of three distinct regimes: (i) $\lambda<1$, where vortices attract each other; (ii) $\lambda=1$, corresponding to the BPS limit with no static forces; and (iii) $\lambda>1$, where vortices repel each other. This classification provides a unique opportunity to investigate how the interplay between internal mode excitation and intersolitonic forces shapes the scattering dynamics of topological defects. Accordingly, the remainder of this section is organized into three subsections, each devoted to a detailed analysis of these regimes. For completeness, we include in the Appendix a collection of figures that allows for a direct global comparison between the different regimes and different values of the excitation amplitude.

The numerical setup employed in this work is based on the schemes introduced in \cite{krusch2024}. The field equations obtained by using the Lorenz gauge condition are discretized. This version of the equations is particularly well suited for the dynamical problems considered here. Spatial derivatives are approximated by fourth-order finite differences, while time evolution is implemented using a second-order accurate scheme. Throughout the simulations, we continuously monitor the fulfillment of the gauge condition, which is satisfied with high precision in all cases reported in this paper. At the boundaries of the computational domain, we impose natural boundary conditions following the prescription described in \cite{krusch2024}. In addition, a damping layer is introduced near the edges of the grid (occupying approximately 6\% of the mesh width) in order to minimize spurious reflections.

Initial configurations are constructed in the standard way by applying Lorentz boosts to static vortex solutions. This procedure requires transforming both the scalar and gauge fields according to the appropriate Lorentz transformation laws. The two vortices are initially placed symmetrically along the $x$-axis at positions $x=\pm d =\pm 9.9$. Since they are well separated at the initial time, the field configuration can be accurately approximated using the Abrikosov ansatz constructed from single 1-vortex solutions. The simulations presented in this work are performed on a square domain $[-30,30]\times [-30,30]$ with spatial resolution $\Delta x= \Delta y= 0.15$. The total simulation time depends on the initial velocity and is typically chosen to be three times the estimated collision time of the vortices. In scenarios where resonant effects lead to multiple-bounce events, this time is extended to up to nine times the collision timescale. The time step is fixed such that the Courant factor is $0.1$, that is, $\Delta t/\Delta x = \Delta t/\Delta y =0.1$, ensuring numerical stability. To construct the velocity and amplitude diagrams, we perform a systematic scan over initial velocities in the range $v_0\in[0.01,0.9]$ with a step size $\Delta v_0=0.001$. In regions where a finer resolution is required to resolve resonant structures, the step is reduced to $\Delta v_0=0.00001$. In addition, the scattering processes are analyzed for initial excitation amplitudes in the interval $\eta_0\in [0.1,1.5]$, with a step $\Delta \eta _0=0.1$. During the evolution, we track the positions of the vortex centers as well as the value of the energy density at their vortex centers. Once the vortices have sufficiently separated and their mutual interaction becomes negligible, these data allow us to extract the final velocities $v_f=(v_x,v_y)$ and the excitation amplitudes $\eta_f$ of the outgoing vortices. The amplitudes are determined using the analytical relation given in \eqref{AmplitudE2}.

We have verified the robustness of our results by performing simulations with different numerical resolutions. In particular, variations in the spatial mesh size were considered, and no significant changes were observed in the behavior of the system. This indicates that the main features of the dynamics are not sensitive to the discretization scheme and, therefore, are not numerical artifacts. We have also analyzed the effect of varying the initial separation between the vortices. In this case, while the overall structure of the velocity and amplitude diagrams remains unchanged, quantitative shifts in the characteristics of the curves are observed. This behavior can be understood in terms of the underlying resonant energy transfer mechanism. As will be discussed below, this mechanism depends on the phases of the vibrational modes of vortices at the instant of collision. Since all simulations are initialized with the same phase, modifying the initial separation effectively alters the collision time and, consequently, the phase of the oscillation at that moment. As a result, changes in the initial vortex positions lead to shifts in the locations of maxima, minima, and the resonant windows within the scattering diagrams. Nevertheless, the global structure of the diagrams, including the oscillatory patterns and fractal features, remains invariant. This confirms that the mechanisms governing vortex scattering in the presence of internal excitations are robust and universal and do not depend on the specific choice of initial conditions.

The behavior described above allows us to propose an approximate analytical expression to construct set of initial velocities where the resonant energy transfer mechanism acts similarly. Consider two vortices initially separated by a distance $2d$ and boosted toward each other with velocity $v_0$. A first approximation for the collision time is then given by $t_c=d/v_0$. It should be emphasized, however, that several effects in the full dynamics reduce the accuracy of this estimate. First, it is well known that the zeros of the complex scalar field, identified with the vortex centers, may exhibit superluminal (ultimately infinite) velocities when two vortex zeros come sufficiently close. This effect implies that the actual collision time is shorter than the simple estimate given above. In addition, inter-vortex interactions may also modify the instantaneous velocity, leading to deviations from uniform motion. To account for these effects, we assume that the phase of the internal excitation at the collision time $t_c = \kappa \cdot d/v_0 $ can be written as
\[
\varphi (v_0) = \kappa \, \frac{d}{v_0} \omega \sqrt{1-v_0^2} + \delta \hspace{0.5cm}, 
\]
where $\kappa$ is a correction factor that effectively incorporates these dynamical deviations, $\omega$ is the oscillation frequency and $\delta$ is a constant phase. Strictly, $\kappa$ is a magnitude depending on the initial separation $d$, the initial velocity $v_0$, the initial amplitude $\eta$ and the coupling constant $\lambda$. 

\begin{figure}[h]
	\centering
	\includegraphics[width=5.cm]{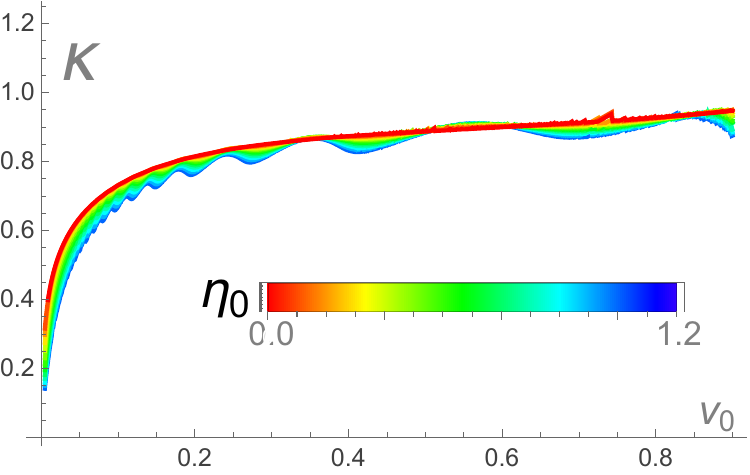} \hspace{0.5cm}\includegraphics[width=5.cm]{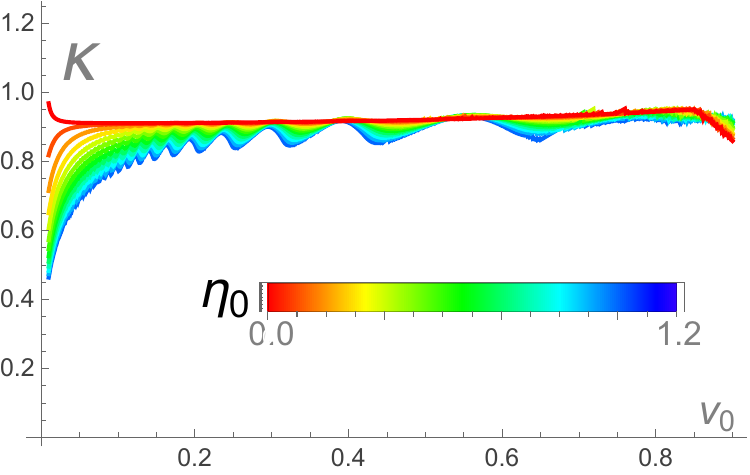} \hspace{0.5cm}\includegraphics[width=5.cm]{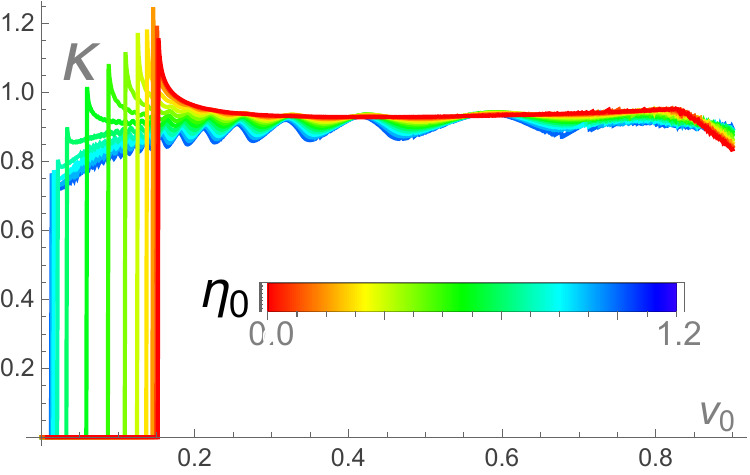} 
	\caption{Collision time correction factor $\kappa$ as a function of the initial velocity and the excitation amplitudes for $\lambda=0.5$ (left), $\lambda=1$ (middle) and $\lambda=1.2$ (right). } \label{fig:ConstanteK}
\end{figure}

In Fig. \ref{fig:ConstanteK} we display the collision time correction factor $\kappa$ as a function of the initial velocity $v_0$ for different values of the initial excitation amplitude in the range $\eta_0\in [0,1.2]$ for the values $\lambda=0.5$, $\lambda=1$ and $\lambda=1.2$, considered as representative of the different regimes of the model. As anticipated, for the case $\lambda=1$, the parameter $\kappa$ is found to be slightly smaller than 1, indicating that the collision between vortices occurs earlier than predicted by the uniform-motion estimate. In the unexcited case, $\eta_0=0$, the collision time remains essentially constant across the range of initial velocities. However, as the initial excitation amplitude increases, small deviations from this constant behavior begin to emerge. These deviations become significantly more pronounced in the low-velocity regime, $v_0<0.1$, where the collision occurs substantially earlier than expected. This behavior can be understood in terms of the effective forces induced by the excitation of internal modes. In particular, when $\eta_0\neq 0$, the spectral flow of the vibrational modes generates an attractive interaction between the vortices. This effect becomes especially relevant at low velocities, where the vortices spend a longer time approaching each other, allowing the attractive force to act more efficiently. As a result, the vortices accelerate during their approach, leading to a noticeable reduction in the collision time compared to the uniform-motion prediction. For $\lambda<1$, the behavior is qualitatively similar to that described above. However, in this regime unexcited vortices already experience attractive intersoliton forces, reducing the value of $\kappa$ even more, particularly at low collision velocities. The case $\lambda>1$, on the other hand, exhibits a qualitatively different behavior due to the repulsive intersoliton forces. In this regime, there exists a minimum initial velocity $v_m$ required for the vortices to overcome the repulsive interaction and eventually collide, see Figure \ref{fig:ConstanteK}(right). This critical velocity is indicated by the vertical line and depends on the initial excitation amplitude. Indeed, the excitation induces an additional attractive interaction that opposes the static repulsive force. Consequently, one expects the critical velocity $v_m$ to decrease as the excitation amplitude increases, in agreement with the behavior shown in Figure \ref{fig:ConstanteK}(right). Motivated by the previous discussion, it is a reasonable approximation to treat the correction factor $\kappa$ as a constant for each value of $\lambda$, at least for sufficiently large initial velocities.

Therefore, treating the initial velocity as a continuous parameter, it is natural to express the phase as a function of $v$, $\varphi (v) = \kappa\, \frac{d}{v} \omega \sqrt{1-v^2} + \delta$. Since the scattering dynamics is governed by the phase of the internal oscillation at the moment of collision, any observable that depends solely on this phase is expected to exhibit a periodic behavior based on the relation
\[
\varphi(v) - \varphi(v_0) = 2\pi j \hspace{0.5cm}, \hspace{0.5cm} j \in \mathbb{Z} \hspace{0.5cm}.
\]
From this periodicity condition, one can infer that a discrete set of initial velocities,
\begin{equation}
	V_j(v_0) = \frac{v_0 \, d \, \omega}{\sqrt{\frac{j^2 4 \pi^2 v_0^2}{\kappa^2} + \frac{4\pi}{\kappa} j v_0 d \omega \sqrt{1-v_0^2} + d^2 \omega^2 }} \hspace{0.5cm}, \hspace{0.5cm} j \in \mathbb{Z} \hspace{0.5cm},  \label{velocidadpatron}
\end{equation}
should display similar dynamical features.

In this work, we consider not only collision processes involving weakly but also strongly excited 1-vortices with large vibrational amplitudes in the range $\eta_0>0.5$, where more extreme behaviors take place. In this regime, it is important to note that the internal modes undergo significant decay over time, see \cite{alonso2024e, arodz1996}. As a consequence, for scattering events with low initial velocities, the interaction time may become sufficiently long that the excitation amplitude at the moment of collision is appreciably smaller than its initial value $\eta_0$. Despite this fact, throughout our analysis we shall continue to take the initial excitation amplitude $\eta_0$ as a reference in our data, rather than attempting to measure the amplitude immediately prior to the collision. The latter procedure would introduce additional uncertainties due to the continuous decay of the vibrational modes during the evolution. In practice, adopting such a definition would mainly result in a rescaling of the scattering observables, leading to more pronounced features in the corresponding diagrams. However, to facilitate the interpretation of our results, the scattering diagrams presented in the Appendix include a dashed gray line indicating the value of the initial excitation amplitude $\eta_0$. In the low-velocity and low-amplitude regime, where vortices are weakly excited and the dynamics can be regarded as approximately adiabatic, one expects the final excitation amplitudes to remain close to this reference value. This behavior is indeed observed in our numerical results for $\eta_0<0.5$.

\subsection{Vortex scattering in the Type I regime $\lambda<1$}

We initiate this study with the analysis of the scattering of two 1-vortices in the Type-I regime of the Abelian Higgs model, where the interaction between vortices is attractive. This case can be regarded as the natural two-dimensional analogue of kink-antikink scattering in the $\phi^4$-model, where kink and antikink also experience a purely attractive force. On these grounds, one may anticipate qualitative similarities between both systems, particularly in relation to resonant energy exchange and multi-bounce phenomena. To illustrate the main features of the dynamics in this regime, we focus on the representative case $\lambda=0.5$, for which the scattering diagrams clearly exhibit the characteristic behavior of Type-I vortices. In this setting, the single-vortex solution supports a single Derrick-type vibrational mode with eigenvalue $\omega^2 = 0.427842$. In order to apply the analytical expressions introduced in \eqref{Seriefyb} and \eqref{DerrickApproximation}, it is necessary to characterize the near-core behavior of both the vortex solution and its associated vibrational mode. This information is encoded in a set of coefficients describing the radial profiles in the vicinity of the vortex center, $\overline{d}_0 = 0.476908$, $\overline{c}_0=0.204048$, $\overline{v}_0 = 0.0657138$ and $\overline{u}_0=-0.121339$.

\begin{figure}[h]
	\centering
	\includegraphics[width=4.cm]{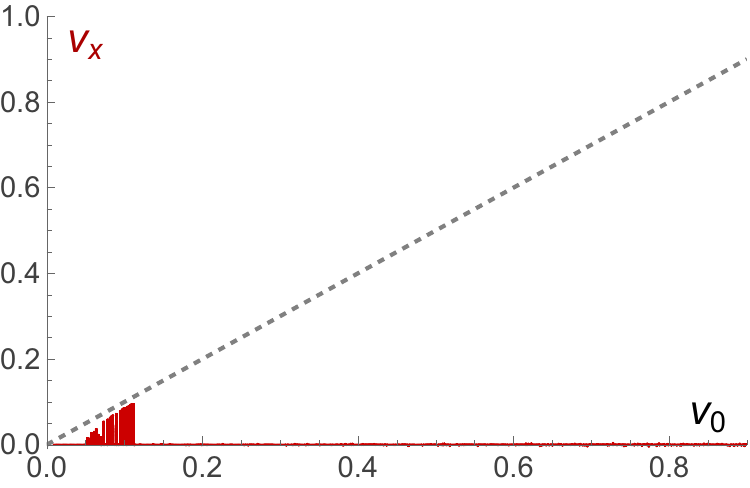} \hspace{0.0cm} \includegraphics[width=4.cm]{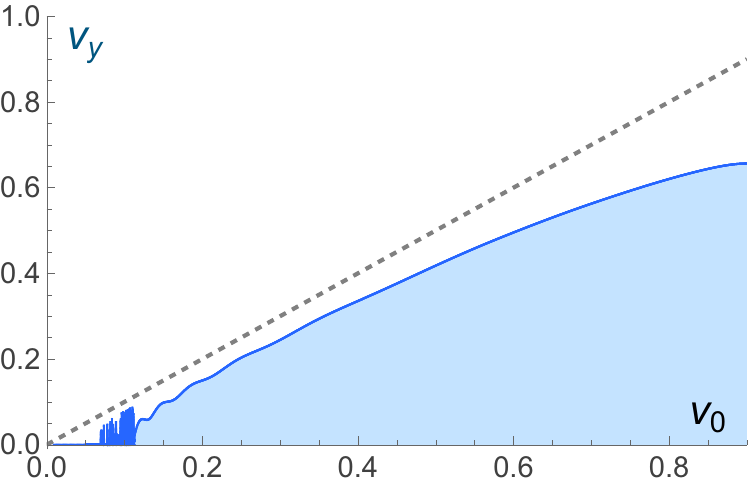} \hspace{0.0cm} \includegraphics[width=4.cm]{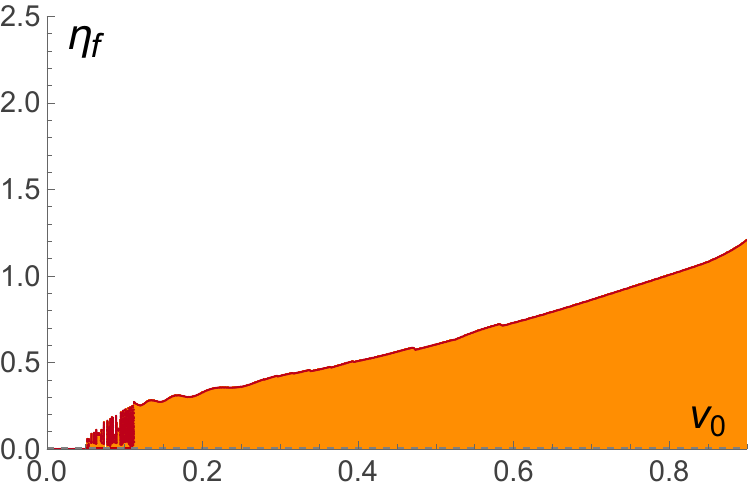} \hspace{0.0cm} \includegraphics[width=4.cm]{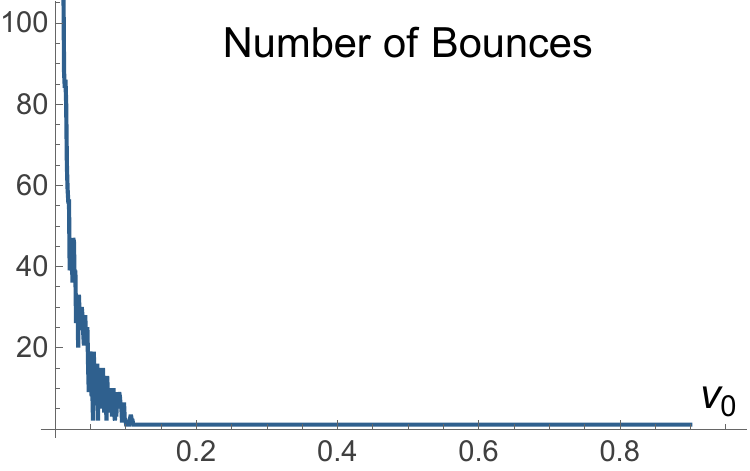}
	\caption{Velocity diagrams for the $v_x$-component (a) and the $v_y$-component (b), together with the amplitude diagram (c) and the number of bounces (d), for head-on scattering of two initially unexcited 1-vortices with $\lambda=0.5$. }
	\label{fig:diagramsL05A00}
\end{figure}

As a first step, we consider the scattering of vortices without initial excitation, $\eta_0=0$. The corresponding scattering diagrams are displayed in Fig. \ref{fig:diagramsL05A00}. More precisely, the first panel displays the $x$-component of the final vortex velocity $v_f$ as a function of the initial collision velocity $v_0$, whereas the second panel shows the corresponding $y$-component. The third panel presents the amplitude scattering diagram, in which the final excitation amplitude $\eta_f$ is plotted as a function of $v_0$. Finally, the fourth panel depicts the number of bounces observed during the simulation as a function of the initial velocity.

A prominent feature of this regime is the existence of a critical velocity $v_c$, associated with the transition between 1-bounce and multi-bounce scattering processes, in close analogy with kink-antikink collisions in the $\phi^4$-model. In the present case, this critical velocity takes the value $v_c=0.11213$. For initial velocities above $v_c$, the vortices undergo a single collision and separate, whereas for $v_0<v_c$ a resonance structure emerges, characterized by a sequence of increasingly narrow multi-bounce windows. This pattern is clearly visible in Fig. \ref{fig:diagramsL05A00}(d), where the number of bounces is plotted as a function of the initial velocity.

\begin{figure}[h]
	\centering
	\includegraphics[width=4.cm]{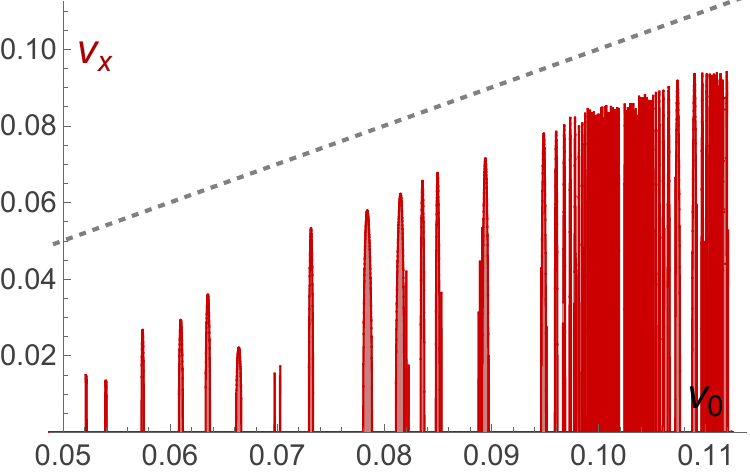} \hspace{0.0cm} \includegraphics[width=4.cm]{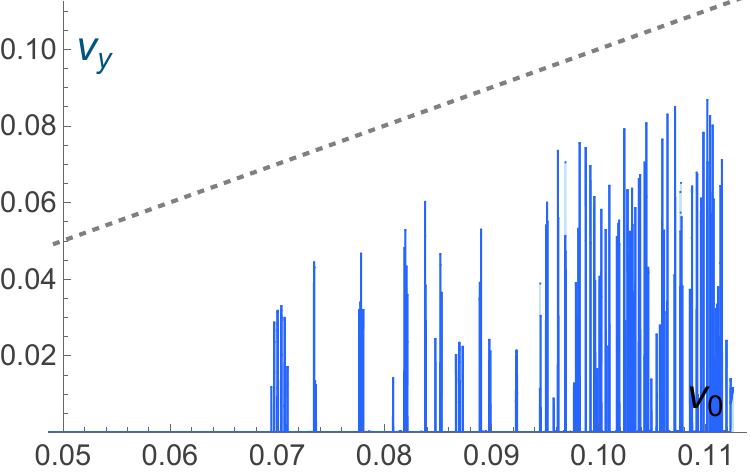} \hspace{0.0cm} \includegraphics[width=4.cm]{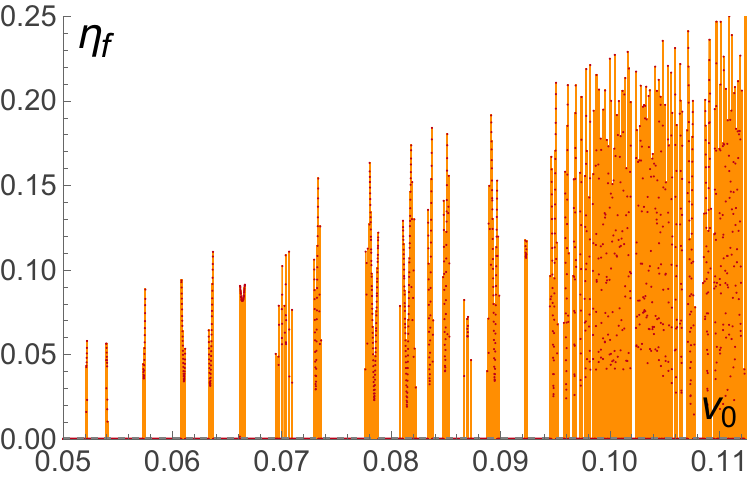} \hspace{0.0cm} \includegraphics[width=4.cm]{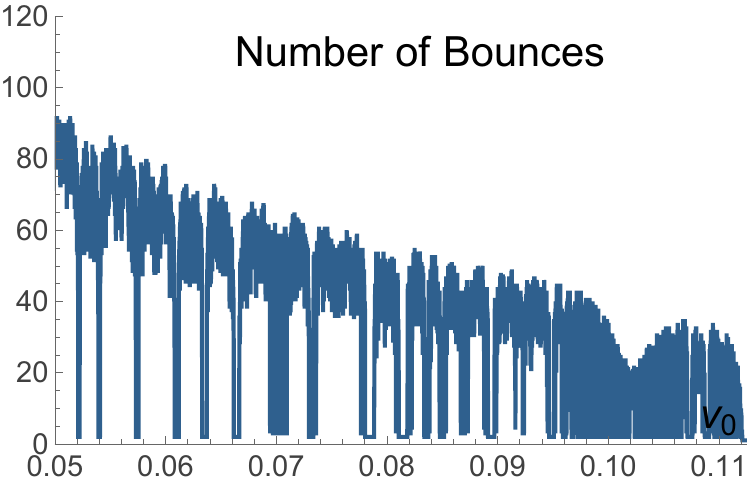}
	\caption{High-resolution scattering diagrams in the interval $v_0\in[0.05,0.11254]$, showing the $v_x$-component (a), the $v_y$-component (b), the final excitation amplitude (c), and the number of bounces (d), for head-on scattering of two initially unexcited 1-vortices with $\lambda=0.5$.}
	\label{fig:diagramsL05A00v05010}
\end{figure}

\begin{figure}[h]
	\centering
	\includegraphics[width=4.cm]{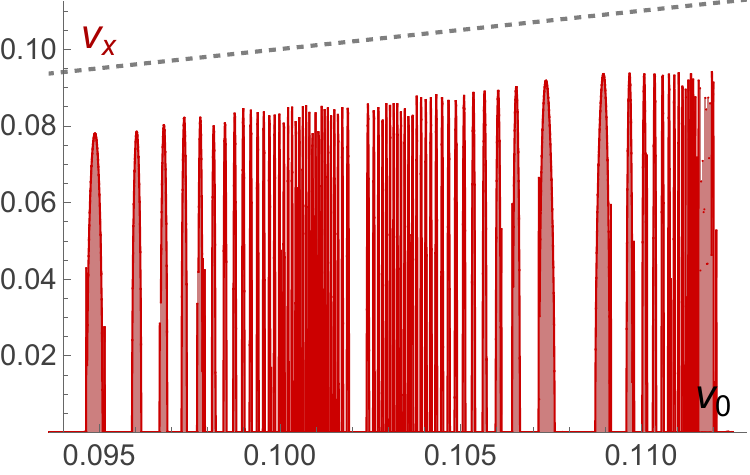} \hspace{0.0cm} \includegraphics[width=4.cm]{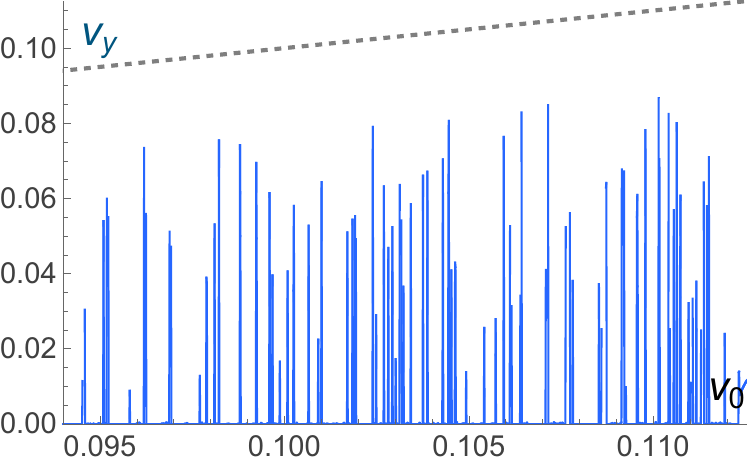} \hspace{0.0cm} \includegraphics[width=4.cm]{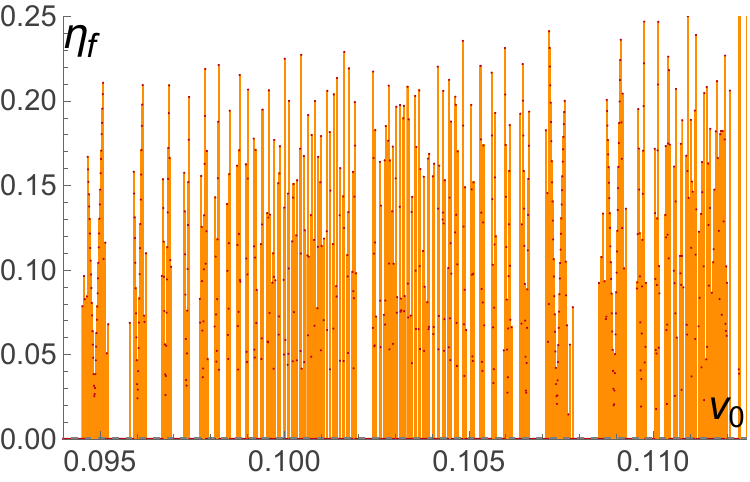} \hspace{0.0cm} \includegraphics[width=4.cm]{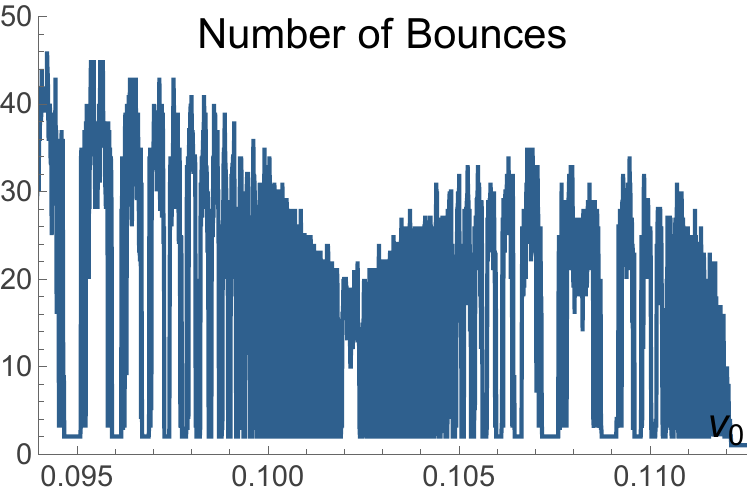}
	\caption{High-resolution scattering diagrams in the interval $[0.094, 0.11254]$, showing the $v_x$-component (a), the $v_y$-component (b), the final excitation amplitude (c), and the number of bounces (d), for head-on scattering of two initially unexcited 1-vortices with $\lambda=0.5$. }
	\label{fig:diagramsL05A00v09010}
\end{figure}

In Figs. \ref{fig:diagramsL05A00v05010} and \ref{fig:diagramsL05A00v09010}, the resolution of the scattering diagrams has been enhanced in order to reveal the aforementioned resonant structure. In these figures, the structure of multi-bounce events, in which vortices undergo several 90° collisions and subsequently separate, is clearly visible, interspersed with episodes of long-lived bound-state formation. This sequence of events is intrinsically linked to the alternation of scattering phenomena in which the resonant energy exchange mechanism is more or less efficient. Furthermore, in Fig. \ref{fig:diagramsL05A00v09010}(d), one can observe the self-similar pattern followed by the alternation of multi-bounce windows and the formation of bound states, which becomes increasingly intricate and narrower as the initial velocity rises until it reaches $v_c$.

A characteristic feature of the non-excited vortex scattering within the type I regime is the absence of elastic scattering events. This is illustrated in the velocity diagrams by the dashed line representing the ideal elastic case, where the final velocity equals the initial one. At low velocities, the dynamics are dominated by the attractive interaction between vortices, which favors the formation of long-lived bound states, whereas higher initial velocities enable the activation of resonant energy exchange mechanisms. This behavior closely parallels that observed in kink–antikink scattering and highlights the crucial role played by attractive forces in shaping the structure of the scattering diagrams.

\begin{figure}[h]
	\centering
	\includegraphics[width=4.cm]{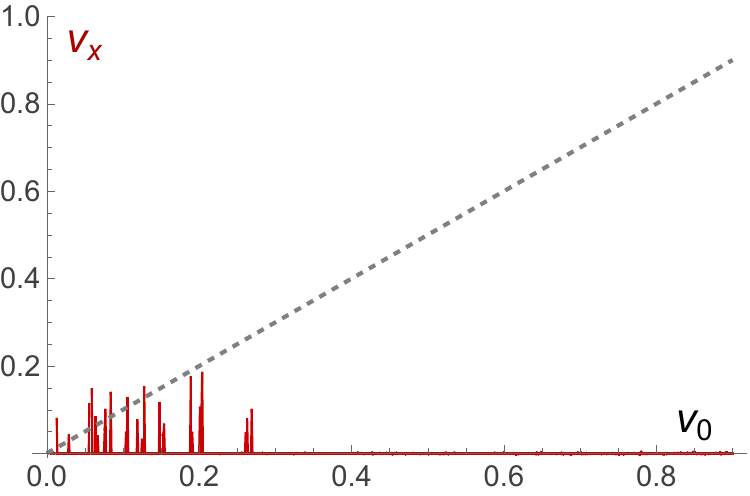} \hspace{0.0cm} \includegraphics[width=4.cm]{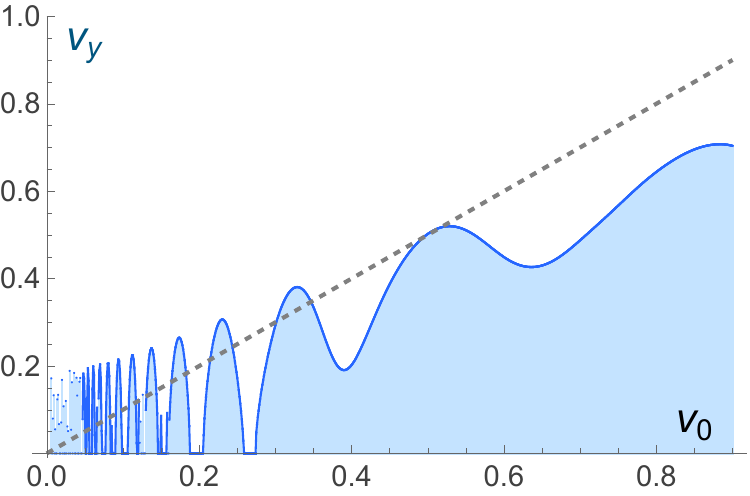} \hspace{0.0cm} \includegraphics[width=4.cm]{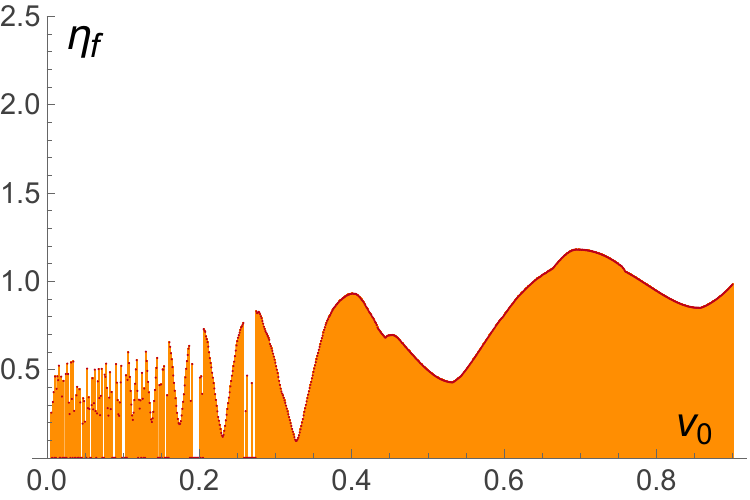} \hspace{0.0cm} \includegraphics[width=4.cm]{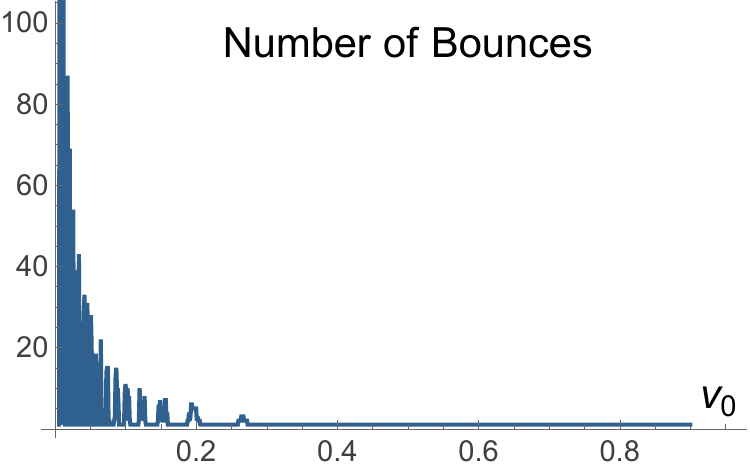}
	\caption{Velocity diagrams for the $v_x$-component (a) and the $v_y$-component (b), together with the amplitude diagram (c) and the number of bounces (d), for head-on scattering of two initially excited 1-vortices with $\lambda=0.5$ and $\eta_0=0.6$.}
	\label{fig:diagramsL05A06}
\end{figure}

As in the kink–antikink system, increasing the initial excitation amplitude $\eta_0$ enhances the resonant energy transfer mechanism, leading to a more chaotic behavior that can be clearly observed in the scattering diagrams. This is illustrated in Fig. \ref{fig:diagramsL05A06} for the case $\eta_0 = 0.6$. In particular, the continuous one-bounce branch observed in Fig. \ref{fig:diagramsL05A00}(c) for initial velocities $v_0 > v_c$ splits into a sequence of isolated one-bounce windows. The gaps between these windows are subsequently filled by new resonance windows, in which the vortices undergo an increasing number of collisions before eventually separating, displaying patterns similar to those shown in Figs. \ref{fig:diagramsL05A00v05010} and \ref{fig:diagramsL05A00v09010}. Overall, the resonance structure becomes progressively more intricate as the initial excitation amplitude increases, as can be seen from the sequence of figures in the left column of the Appendix. This dramatic increase in the complexity of the scattering process appears to rely crucially on the presence of attractive intersoliton forces, in close analogy with kink–antikink collisions in the $\phi^4$-model.

In this regime, we can check the equation (\ref{velocidadpatron}) in order to predict the values of the initial velocities for which the resonant energy transfer mechanism is most efficient, namely those corresponding to the maxima of the final velocity observed in Figure 5. In Table 1, we list the values of the initial velocities associated to the maxima of the function $v_f$ (denoted as $\arg \max v_f$) and compare them with the values predicted by (27), using as reference the last peak situated at $v_0 \approx 0.883$ and the correction factor $\kappa \approx 0.787$. We find an good agreement between the numerical results (second column) and the analytical predictions (fourth column).

\begin{table}[h]
	\centering
	\begin{tabular}{|c||c|c||c|} \hline
		$j$ & $\arg \max v_f$ & $\max v_f$ & $V_j(0.883)$ \\ \hline\hline 
		0 & 0.883 & 0.707 & 0.883  \\
		1 & 0.529 & 0.520 &  0.493  \\
		2 & 0.329 & 0.381 &  0.316  \\
		3 & 0.230 & 0.306 &  0.230  \\
		4 & 0.174 & 0.265 &  0.180  \\
		5 & 0.138 & 0.240 &  0.148  \\
		6 & 0.112 & 0.225 &  0.125  \\
		7 & 0.094 & 0.215 &  0.108  \\
		$\vdots$ & $\vdots$ & $\vdots$ & $\vdots$  \\ \hline 
	\end{tabular} 
	\caption{Initial velocities associated with the resonance peaks in the scattering diagram with $\lambda=0.5$ and $\eta_0=0.6$ (see Fig. \ref{fig:diagramsL05A06}(b)), together with the corresponding maximum final velocities and the theoretical predictions obtained from \eqref{velocidadpatron}. } \label{table0} 
\end{table}

Both Table \ref{table0}  and Fig. \ref{fig:diagramsL05A06} reveal the presence of superelastic scattering events, in which the vortices emerge from the collision with a final velocity greater than their initial one as a result of extracting part of the energy stored in their internal vibrational modes. Another important feature that can be inferred from expression (\ref{velocidadpatron}) is the self-similar replication of the velocity 1-bounce windows as the initial velocity decreases. In particular, the pattern repeat at progressively smaller velocity scales with smaller amplitudes, giving rise to a fractal-like structure in the scattering diagrams. This behavior can ultimately be traced back to the relativistic nature of the model: the dependence of the phase on the Lorentz factor introduces a nontrivial scaling with the velocity, which underlies the emergence of this self-similar structure.

\subsection{Vortex scattering in the self-dual regime $\lambda=1$}

We next consider the self-dual regime, $\lambda=1$, where the static interaction between vortices vanishes. In contrast to the attractive regime $\lambda<1$, this case offers an ideal framework for isolating the effects of the internal vibrational modes on the scattering dynamics, free from the influence of additional intersoliton forces. For this reason, we now investigate this regime in greater detail. The radial profiles of the vortex near the origin are characterized by the coefficients $\overline{d}_0 = 0.603005$ and $\overline{c}_0=0.25$, see \eqref{Seriefyb}. On the other hand, the Derrick-type mode has an eigenvalue $\omega^2 = 0.777456$ and, furthermore, the behavior of the corresponding eigenfunctions for $r\approx 0$ is determined by the coefficients $\overline{v}_0 = 0.129188$ and $\overline{u}_0=-0.200486$, see \eqref{DerrickApproximation}, \cite{goodband1995, alonso2016a, alonso2016b}. 

\begin{figure}[h]
	\centering
	\includegraphics[width=4.cm]{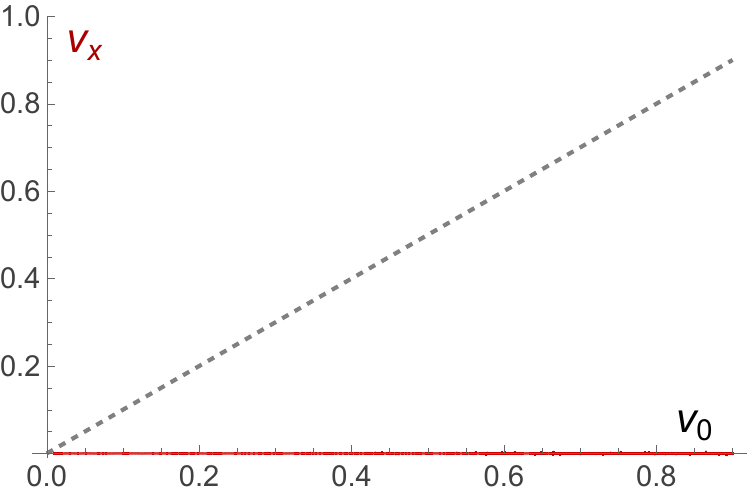} \hspace{0.0cm} \includegraphics[width=4.cm]{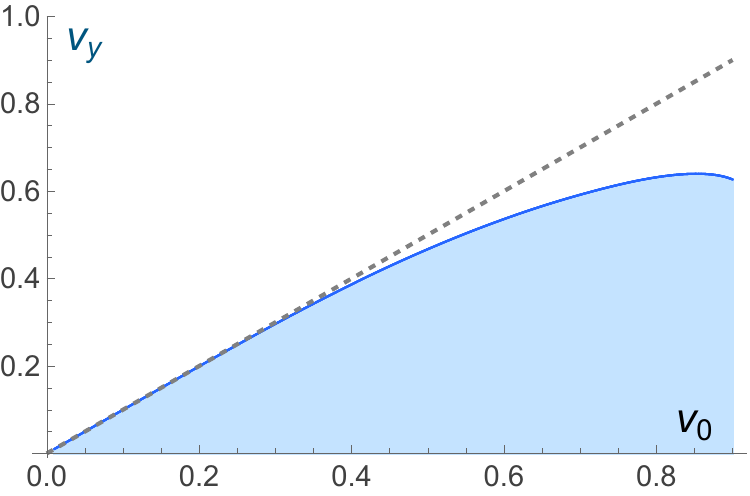} \hspace{0.0cm} \includegraphics[width=4.cm]{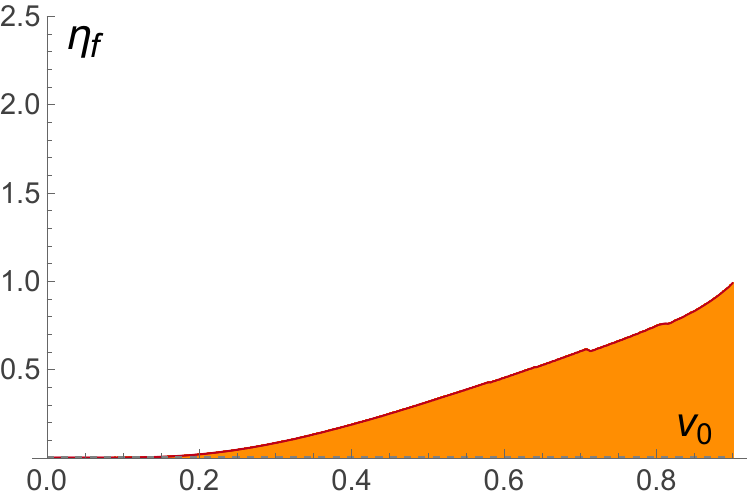} \hspace{0.0cm} \includegraphics[width=4.cm]{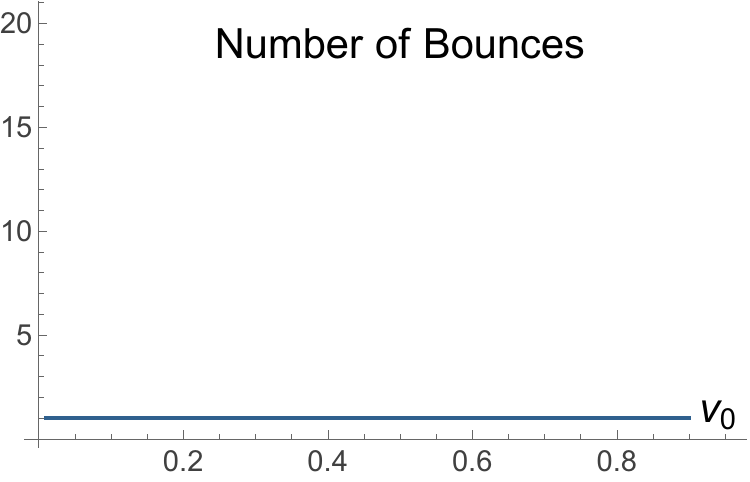}
	\caption{Velocity diagrams for the $v_x$-component (a) and the $v_y$-component (b), together with the amplitude diagram (c) and the number of bounces (d), for head-on scattering of two initially unexcited 1-vortices with $\lambda=1$.} \label{fig:diagramsL10A00}
\end{figure}

We first describe the global features of the scattering process in this scenario. As a reference case, Fig. \ref{fig:diagramsL10A00} presents the velocity and amplitude diagrams for collisions of unexcited vortices. In this situation, vortices always collide only once (see Fig. \ref{fig:diagramsL10A00}(d)) at the origin of the plane for any value of the initial velocity and subsequently $90^\circ$-scatter along the $y$-axis with a certain final velocity $v_f=(0,v_y)$. For low initial velocities, when the scattering phenomena can be described in terms of the moduli space approximation, the process is effectively elastic. The numerical results, shown as a blue curve, closely follow the elastic dashed line in the low-velocity regime. However, as the collision velocity increases, the curve progressively deviates from the elastic line, indicating the onset of inelastic effects. In this regime, part of the energy is radiated away and part is transferred into internal excitations of the vortices. This is further confirmed by the corresponding amplitude diagram, see Fig. \ref{fig:diagramsL10A00}(c), where a nonzero excitation amplitude is observed after the collision.

\begin{figure}[h]
	\centering
	\includegraphics[width=4.cm]{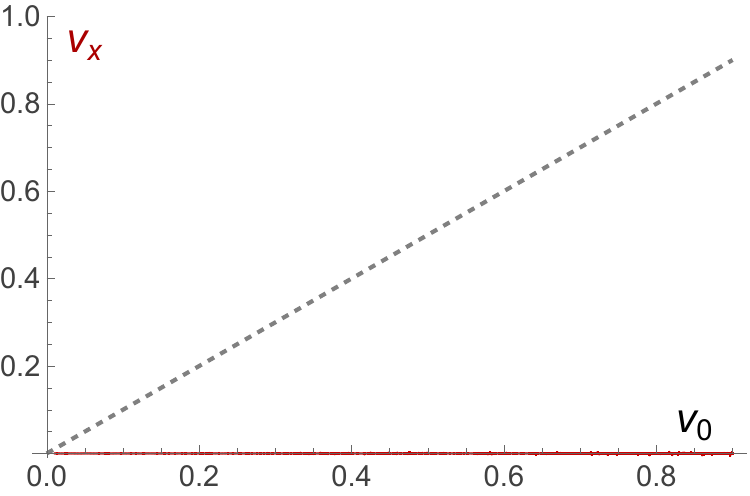} \hspace{0.0cm} \includegraphics[width=4.cm]{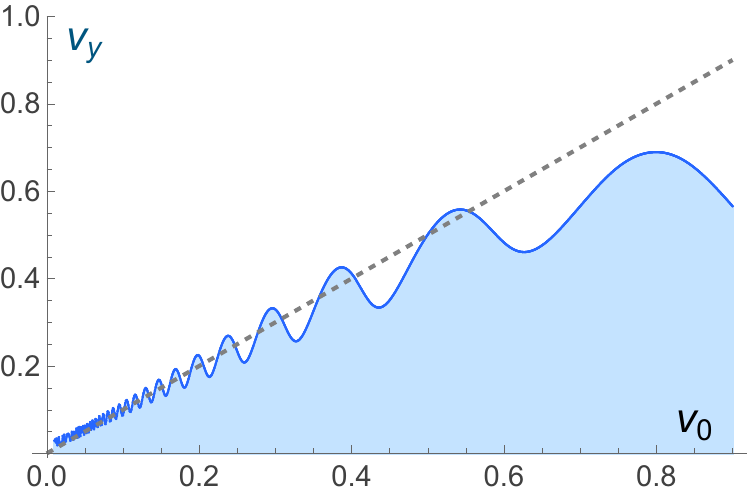} \hspace{0.0cm} \includegraphics[width=4.cm]{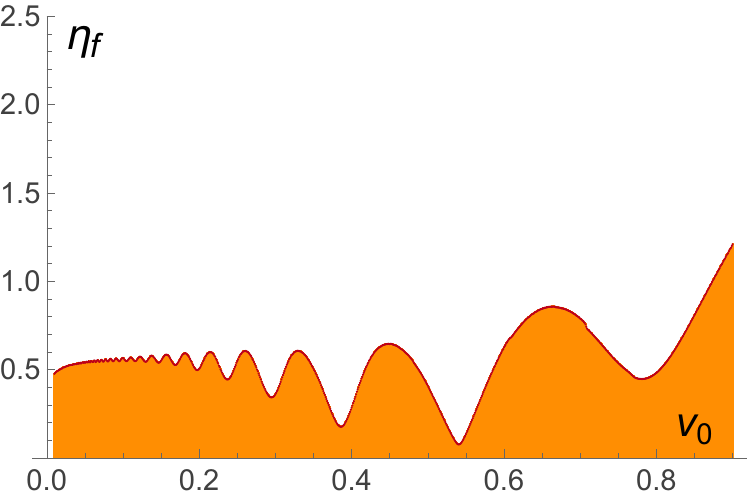} \hspace{0.0cm} \includegraphics[width=4.cm]{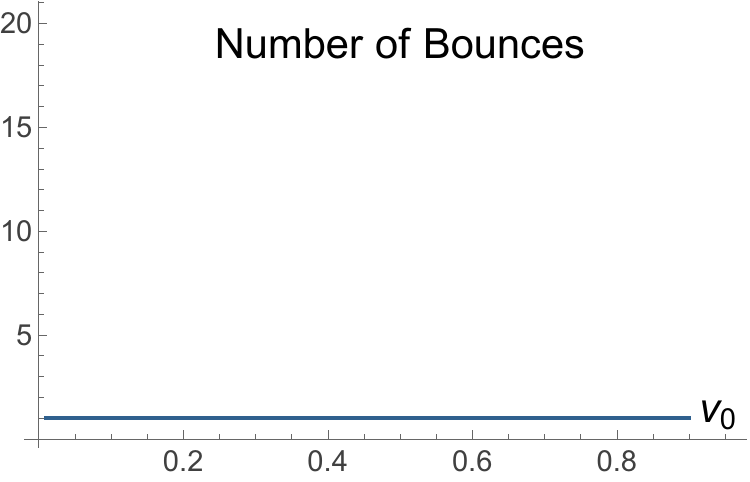}
	\caption{Velocity diagrams for the $v_x$-component (a) and the $v_y$-component (b), together with the amplitude diagram (c) and the number of bounces (d), for head-on scattering of two initially excited 1-vortices with $\lambda=1$ and $\eta_0=0.6$.}
	\label{fig:diagramsL10A06}
\end{figure}

For comparison, we next consider the case in which the vortices are initially excited with a relatively large amplitude, $\eta_0=0.6$, see Fig. \ref{fig:diagramsL10A06}. In this scenario, the scattering dynamics becomes significantly more intricate. In particular, the resonant energy transfer mechanism can be activated even at relatively low collision velocities, resulting in a highly nontrivial exchange between vibrational and translational degrees of freedom. 

One of the most striking features within this regime is the oscillatory dependence of the final velocity on the initial velocity. The velocity diagram is no longer monotonic but instead exhibits a sequence of local maxima and minima. These oscillations reflect the alternating dominance of energy transfer channels: local maxima correspond to events where vibrational energy is efficiently converted into kinetic energy, leading to final velocities exceeding the initial ones (super-elastic scattering), whereas local minima indicate the opposite process, where kinetic energy is transferred into internal modes. This behavior is clearly observed in the amplitude diagram. A direct comparison between the two diagrams reveals a strong correlation: maxima in the velocity diagram correspond to minima in the amplitude diagram, and vice-versa.

\begin{table}[h]
	\centering
	\begin{tabular}{|c||c|c||c|c||c|c|} \hline
		$j$ & $\arg \max v_f$ & $\arg \min \eta_f$& $\arg \min v_f$ & $\arg \max \eta_f$ & $V_j(0.800)$ & $V_j(0.627)$ \\ \hline\hline 
		0 & 0.800 & 0.781 & 0.627 & 0.663 & 0.800 & 0.627 \\
		1 & 0.543 & 0.541 & 0.436 & 0.448 & 0.537 & 0.436 \\
		2 & 0.387 & 0.386 & 0.327 & 0.329 & 0.386 & 0.328 \\
		3 & 0.296 & 0.295 & 0.259 & 0.261 & 0.298 & 0.261 \\
		4 & 0.238 & 0.238 & 0.213 & 0.214 & 0.241 & 0.216 \\
		5 & 0.198 & 0.198 & 0.181 & 0.182 & 0.202 & 0.184 \\
		6 & 0.169 & 0.168 & 0.156 & 0.157 & 0.174 & 0.160 \\
		7 & 0.147 & 0.147 & 0.137 & 0.138 & 0.152 & 0.142 \\
		$\vdots$ & $\vdots$ & $\vdots$ & $\vdots$ & $\vdots$ & $\vdots$ & $\vdots$ \\ \hline 
	\end{tabular} 
	\caption{
    Initial velocities associated with the local extrema in the velocity and amplitude scattering diagrams with $\lambda=1.0$ and $\eta_0=0.6$ (see Fig. \ref{fig:diagramsL10A06}(b)), together with the theoretical predictions obtained from \eqref{velocidadpatron}.  } \label{table1} 
\end{table}

Table \ref{table1} provides the initial velocities $v_0$ associated to maxima and minima observed in the velocity and amplitude diagrams for the case $\eta_0=0.6$, ordered by decreasing values of $v_0$. A clear correlation can be identified: maxima in the final velocity $v_f$ correspond to minima in the final excitation amplitude $\eta_f$, while minima in $v_f$ are associated with maxima in $\eta_f$, as previously mentioned. This correlation provides clear evidence of the underlying resonant energy exchange mechanism. Furthermore, the local maxima and minima observed in the velocity diagrams are expected to correspond to scattering events with similar dynamical characteristics. In particular, their positions should be approximately predicted by the analytical expression derived in \eqref{velocidadpatron}. In Table I, we present the values obtained from this formula, incorporating a corrected estimate for the collision time given by $\kappa=0.879$. The sequence of maxima is generated by taking as a reference the initial velocity $v_0=0.8$, while the minima are obtained by choosing the initial value $v_0=0.627$, from which the remaining sequence follows. A comparison with the numerical data shows that the agreement is remarkably good, with deviations typically of order $10^{-3}$. This level of accuracy supports the validity of the phase-based approximation in capturing the structure of the scattering diagrams.

\begin{figure}[h]
	\centering
	\includegraphics[width=5cm]{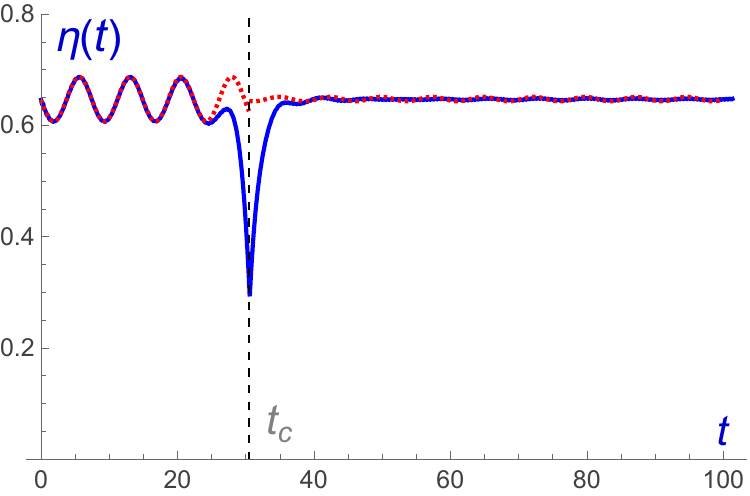} \hspace{0.5cm}\includegraphics[width=5cm]{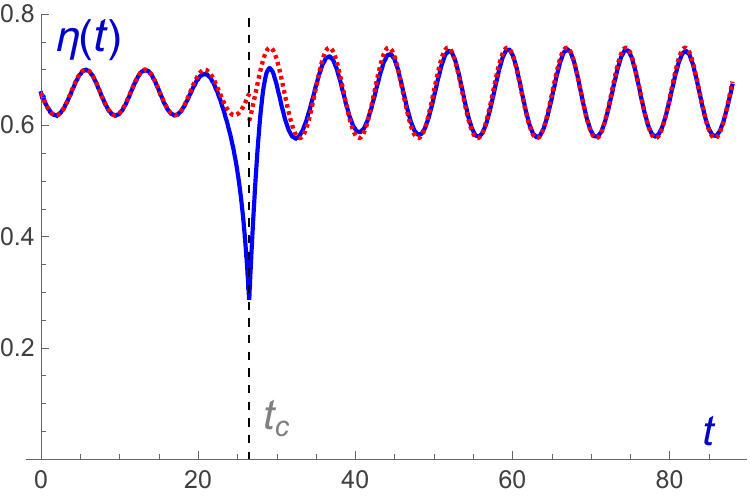} \hspace{0.5cm}  \includegraphics[width=5cm]{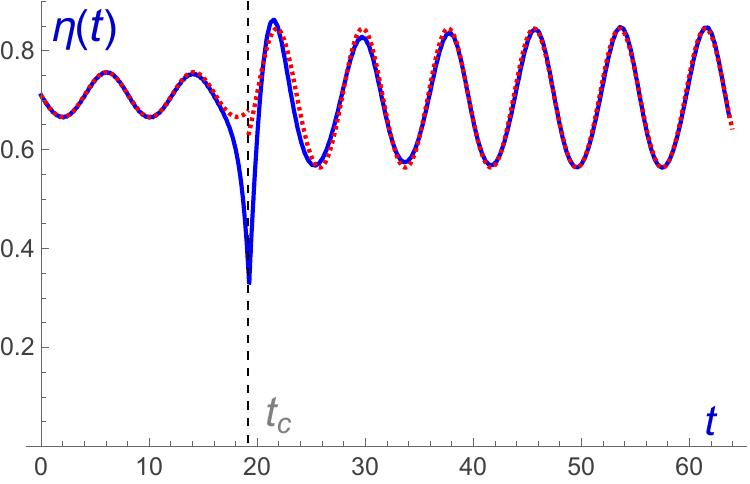}
	\caption{Time evolution of the excitation amplitude for the collision of two 1-vortices with $\lambda=1$, initial amplitude $\eta_0=0.1$ and initial velocities $v_0=0.296$ (left), $v_0=0.341$ (middle) and $v_0=0.471$ (right). $t_c$ denotes the collision time.} \label{fig:AmplitudeEvolution}
\end{figure}

The previous observations support the interpretation that energy transfer is maximized when the internal oscillation reaches either a maximum or a minimum at the instant of collision. To illustrate this mechanism, Fig. \ref{fig:AmplitudeEvolution} shows the time evolution of the energy-density peak for representative scattering events corresponding to a local maximum and a local minimum in the velocity diagram. In Fig. \ref{fig:AmplitudeEvolution} (left), we present the evolution for a collision with initial velocity $v_0=0.296$ and initial excitation amplitude $\eta_0=0.1$. The collision occurs at the time $t_c\approx 30.5$ introduced in the figure. After the interaction, the amplitude decreases to $\eta_f \approx  0.01$, while the final velocity increases to approximately $0.30$, indicating a transfer of energy from the vibrational to the translational sector. Additionally, the analytical prediction given by (\ref{GeneralAmplitude}) with $\eta=\eta_0$ and $\eta=\eta_f$ is shown as red dashed curves before and after the collision, exhibiting excellent agreement with the numerical results. Deviations are only observed during the short collision interval, where the two 1-vortices merge into a transient 2-vortex configuration. In this regime, the energy density develops an annular structure, with a reduced value at the center of the composite configuration. Overall, this example illustrates that when the vortices collide near a minimum of the oscillation cycle, energy is efficiently transferred from the internal vibrational modes to the kinetic motion of the outgoing vortices.

The scattering events shown in Fig. \ref{fig:AmplitudeEvolution} (middle and right) illustrate the complementary process in which energy is transferred from the kinetic to the vibrational sector. In  Fig. \ref{fig:AmplitudeEvolution} (middle), the vortices approach each other with an initial velocity $v_0=0.341$ and the same initial excitation amplitude $\eta_0=0.1$. After the collision at $t_c\approx 26.5$, the final velocity is reduced to $v_f\approx 0.33$, while the excitation amplitude becomes $\eta_f\approx 0.2$, indicating that part of the kinetic energy has been converted into internal vibrational energy. As in the previous example, the analytical prediction (\ref{GeneralAmplitude}), shown as dashed red curves, is in excellent agreement with the numerical evolution outside the interaction region. In this case, the energy transfer occurs when the vortices collide near a maximum of the oscillation cycle, where the vibrational amplitude is largest. In the last case, the initial velocity is greater than in the previous case, $v_0=0.471$ with the same initial amplitude $\eta_0=0.1$. The scattering output is characterised by a decrease in the velocity $v_f\approx 0.4$ and an increase in the vibrational excitation $\eta_f=0.32$

\begin{figure}[h]
	\centering
	\includegraphics[width=4.cm]{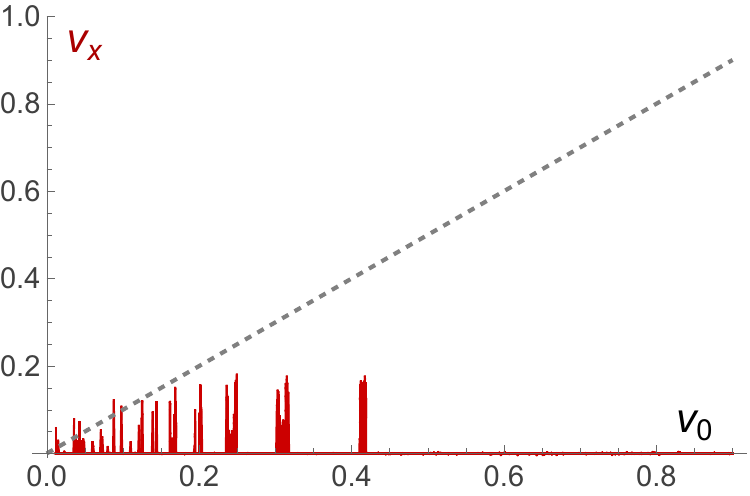} \hspace{0.0cm} \includegraphics[width=4.cm]{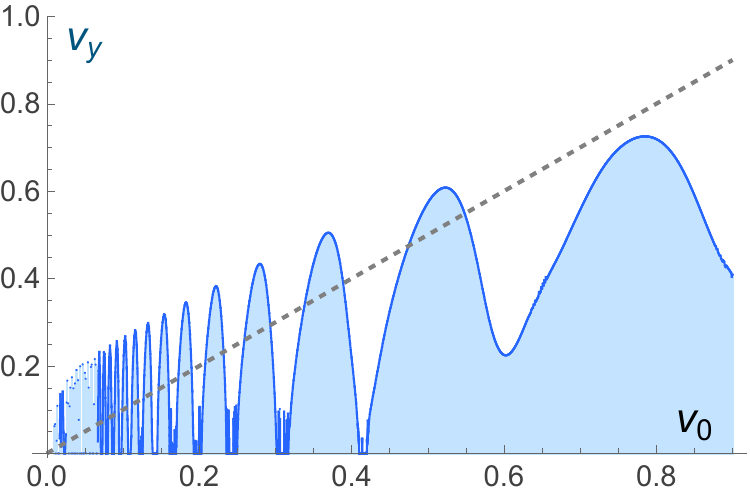} \hspace{0.0cm} \includegraphics[width=4.cm]{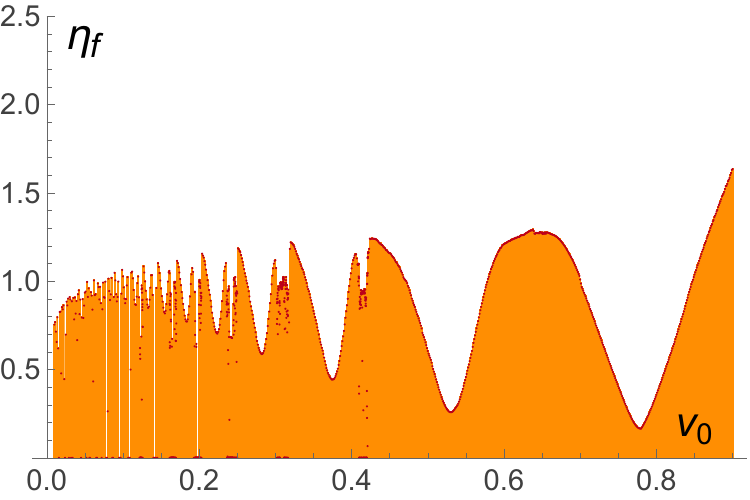} \hspace{0.0cm} \includegraphics[width=4.cm]{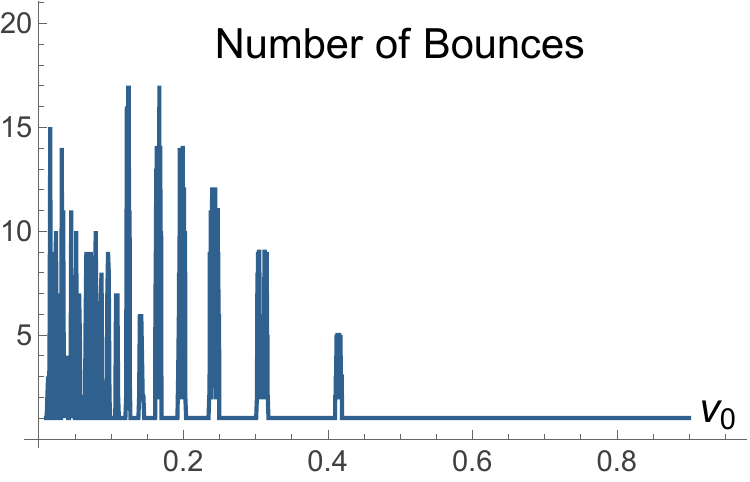}
	\caption{Velocity diagrams for the $v_x$-component (a) and the $v_y$-component (b), together with the amplitude diagram (c) and the number of bounces (d), for head-on scattering of two initially excited 1-vortices with $\lambda=1$ and $\eta_0=1.5$.}
	\label{fig:diagramsL10A15}
\end{figure}

The final case considered in this section corresponds to scattering processes with a large initial excitation amplitude, $\eta_0=1.5$, as shown in Fig. \ref{fig:diagramsL10A15}. In this case, the oscillations of the final velocity as a function of the initial velocity become sufficiently pronounced that their minima reach the axis $v_f=0$, as can also be seen in the sequence of graphics given in the Appendix. As a consequence, the velocity diagram develops a set of isolated velocity windows in which the vortices undergo a single $90^\circ$ scattering event. Between these windows, however, there exist additional intervals associated with multiple-bounce processes. This behavior is clearly confirmed in Fig. \ref{fig:diagramsL10A15}(d), where the number of bounces is displayed as a function of the initial velocity $v_0$. It should be noted that the number of bounces is determined within the finite duration of our simulations. In some cases, the vortices may continue to interact through repeated collisions over longer timescales, eventually releasing enough vibrational energy to separate. 

\begin{figure}[h]
	\centering
	\includegraphics[width=4.cm]{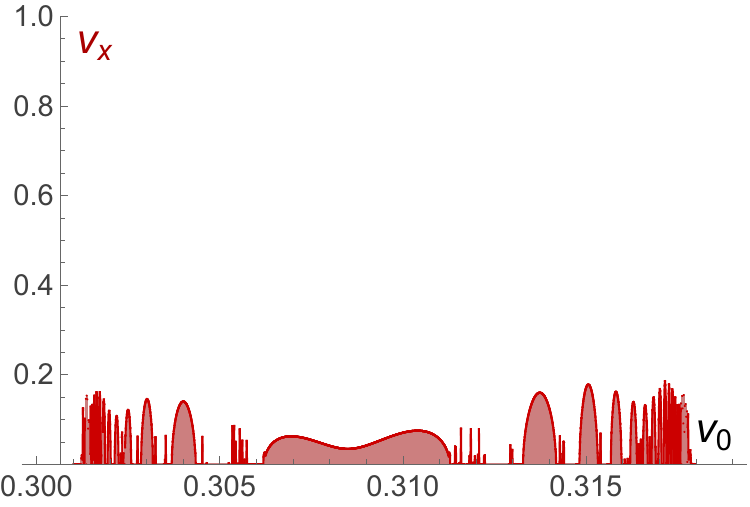} \hspace{0.0cm} \includegraphics[width=4.cm]{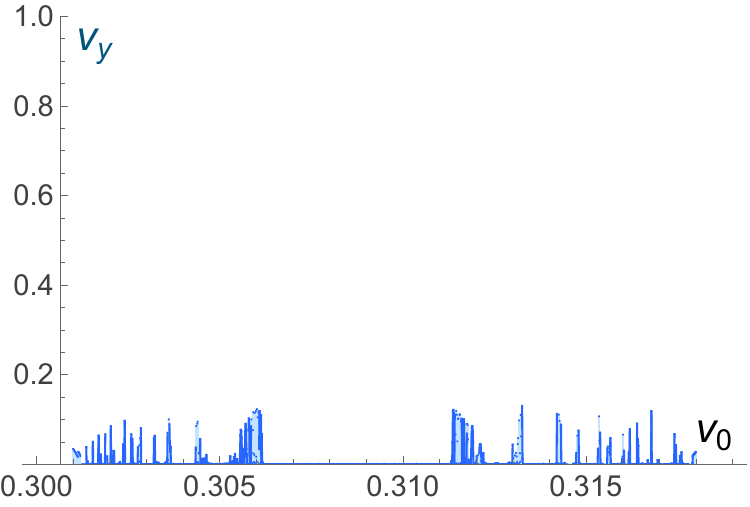} \hspace{0.0cm} \includegraphics[width=4.cm]{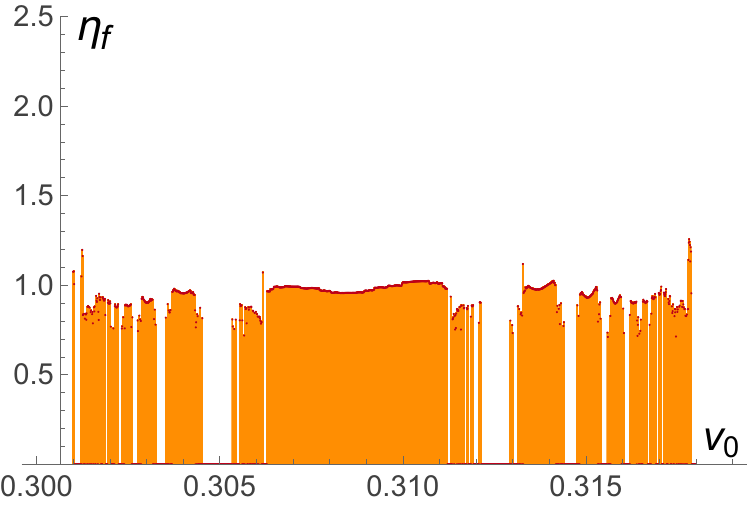}  \hspace{0.0cm} \includegraphics[width=4.cm]{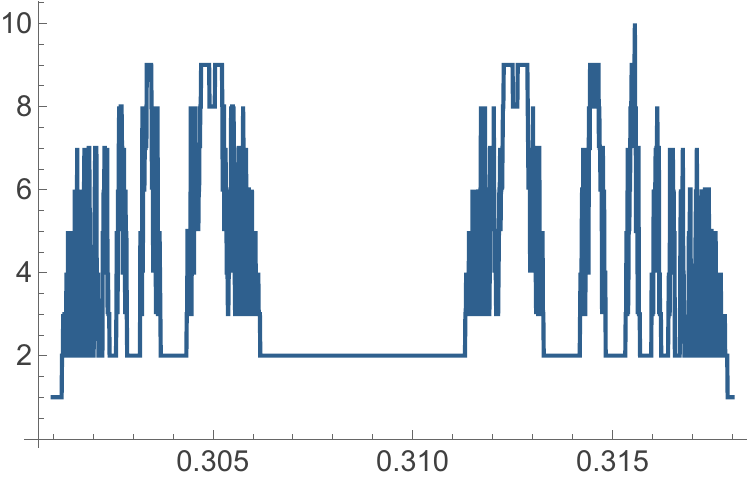}
	\caption{Velocity diagrams for the $v_x$-component (a) and the $v_y$-component (b), together with the amplitude diagram (c) and the number of bounces (d), for head-on scattering of two initially excited 1-vortices with $\lambda=1$ and $\eta_0=1.5$ in the velocity interval $[0.3012,03179]$.}
	\label{fig:diagramsL10A15Ref}
\end{figure}

To gain further insight into the behavior within the intervals separating the 1-bounce windows, we present in Fig. \ref{fig:diagramsL10A15Ref} the scattering diagrams in the range $v_0\in [0.3012,03179]$. This refinement allows for a much clearer visualization of the chaotic dynamics and reveals the underlying fractal structure present in this region. 

It is also noteworthy that there is a striking similarity in the structure of the scattering diagrams shown in Figs. \ref{fig:diagramsL05A06} and \ref{fig:diagramsL10A15}, corresponding to the regimes $\lambda<1$ and $\lambda=1$, respectively. These figures are associated with initial excitation amplitudes $\eta_0=0.6$ and $\eta_0=1.5$, respectively. In other words, in order to reproduce a resonance pattern similar to that observed in Figure \ref{fig:diagramsL05A06} for $\lambda=0.5$ and $\eta_0=0.6$, it is necessary to increase the initial excitation amplitude in the self-dual case $\lambda=1.0$ up to $\eta_0=1.5$. This observation suggests that the attractive force, which arises naturally in the Type I regime, must instead be effectively generated in the self-dual case through the excitation of internal modes. In this sense, increasing the excitation amplitude enhances the effective attractive interaction induced by the vibrational degrees of freedom, thereby reproducing the resonance structure observed in the attractive regime.

\subsection{Vortex scattering in the Type II regime $\lambda>1$}

In this section, we investigate the Type-II regime of the Abelian Higgs model, characterized by repulsive interactions between the 1-vortices. In order to present our results in a concrete setting, we focus on the representative case $\lambda=1.2$. For this value of the coupling constant, the 1-vortex again supports a single Derrick-type internal mode, with eigenvalue $\omega^2 = 0.886831$. The behavior of the vortex profiles near the origin is determined by the coefficients $\overline{d}_0 = 0.642640$ and $\overline{c}_0=0.263343$, as defined in \eqref{Seriefyb}. Similarly, the corresponding Derrick mode is characterized near the vortex center by the coefficients $\overline{v}_0 = 0.148143$ and $\overline{u}_0=-0.212036$, see \eqref{DerrickApproximation}. 

\begin{figure}[h]
	\centering
	\includegraphics[width=4.cm]{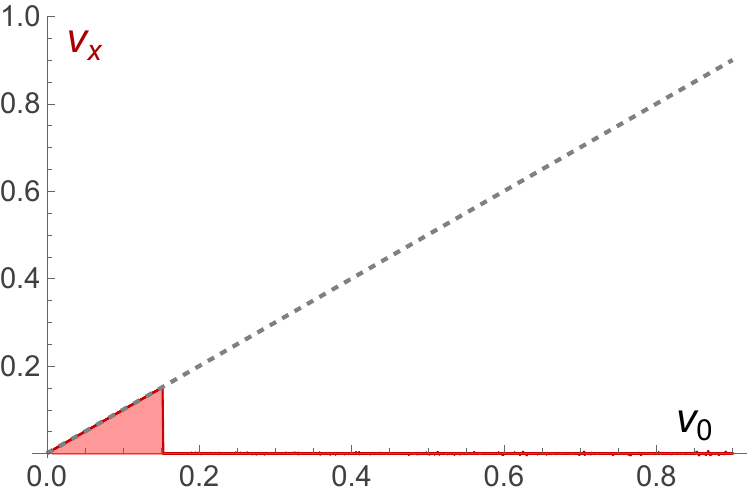} \hspace{0.0cm} \includegraphics[width=4.cm]{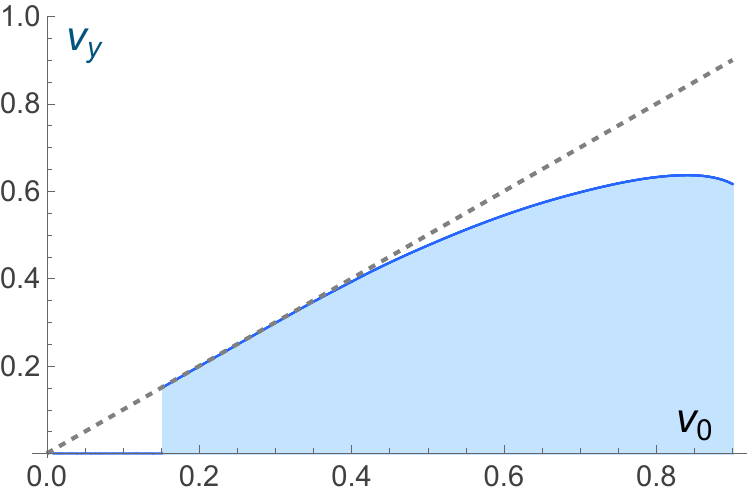} \hspace{0.0cm} \includegraphics[width=4.cm]{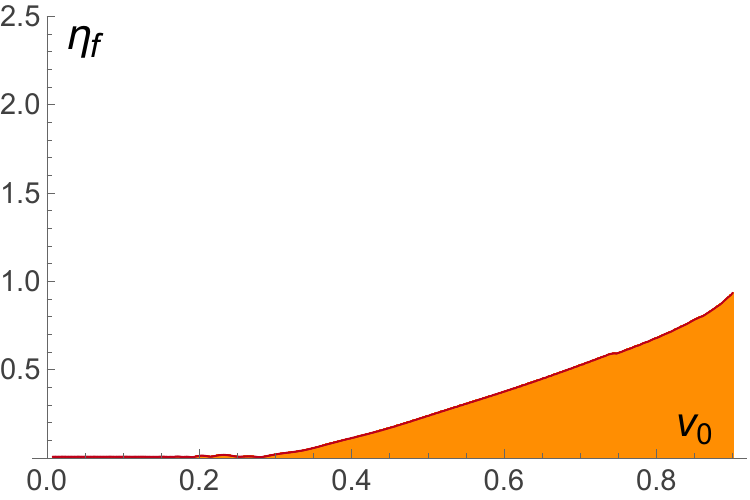} \hspace{0.0cm} \includegraphics[width=4.cm]{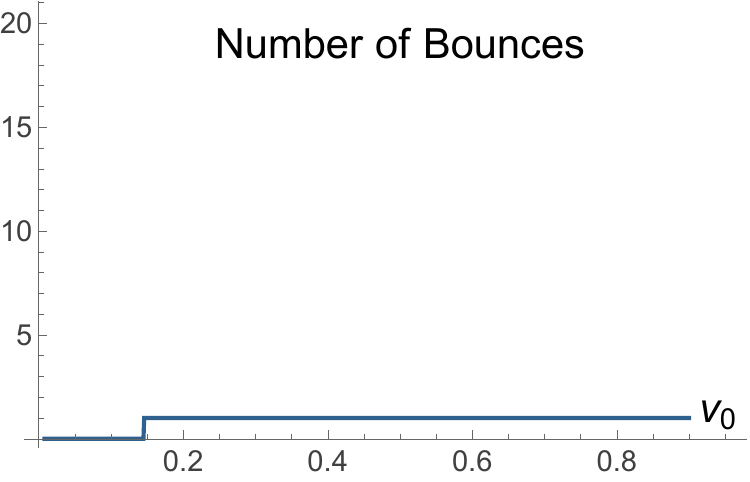}
	\caption{Velocity diagrams for the $v_x$-component (a) and the $v_y$-component (b), together with the amplitude diagram (c) and the number of bounces (d), for head-on scattering of two initially excited 1-vortices with $\lambda=1.2$ and $\eta_0=0.0$.}
	\label{fig:diagramsL12A00}
\end{figure}

We begin by analyzing the scattering of unexcited vortices, $\eta_0=0$. The corresponding scattering diagrams for $\lambda=1.2$ are shown in Fig. \ref{fig:diagramsL12A00}. The first qualitative difference with respect to the self-dual case emerges at low initial velocities. In this case, the repulsive force between vortices dominates over their inertia, preventing collision. As a result, the vortices reverse their motion along the $x$-axis and separate with a final velocity approximately equal to the initial one, indicating that these scattering process are quasi-elastic, see Fig. \ref{fig:diagramsL12A00}(a). Besides, there exists a threshold velocity $v_m=0.153$ above which the kinetic energy overcomes the repulsive interaction, allowing the vortices to collide. For initial velocities exceeding this threshold, the vortices undergo the characteristic $90^\circ$-scattering, emerging along the $y$-axis with a final velocity that, for moderate values of $v_0$, remains close to the initial one. This again corresponds to an approximately elastic regime, see Fig. \ref{fig:diagramsL12A00}(b). At higher collision velocities, nonlinear effects become increasingly important. In particular, part of the kinetic energy is transferred into internal vibrational modes and radiative channels, leading to deviations from elastic behavior. This is reflected in the departure of the final velocity from the reference elastic line (shown as a dashed gray line in the figures), as can be observed in Fig. \ref{fig:diagramsL12A00}(b).

Therefore, the specific qualitative feature of the Type-II regime is the existence of a minimum velocity required for vortex collision. This is clearly illustrated in Fig. \ref{fig:diagramsL12A00}(d), where the number of collisions is plotted as a function of the initial velocity. Moreover, the range of initial velocities exhibiting elastic behavior appears to be broader than in the self-dual case, reflecting the dominant role of repulsive forces at low energies.

\begin{figure}[h]
	\centering
	\includegraphics[width=4.cm]{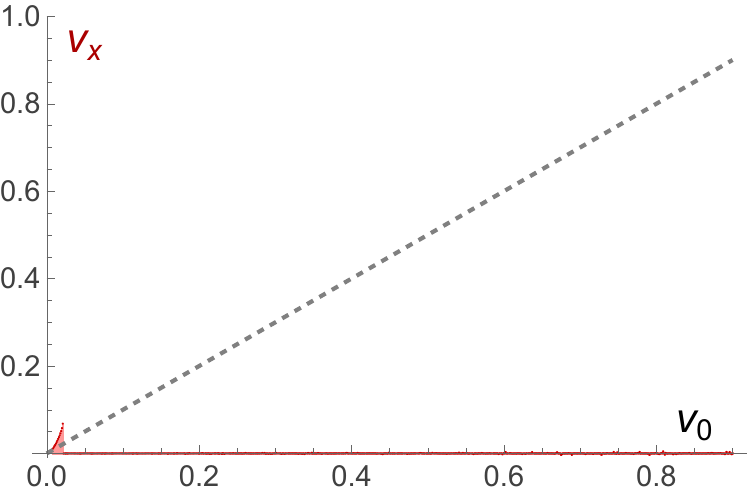} \hspace{0.0cm} \includegraphics[width=4.cm]{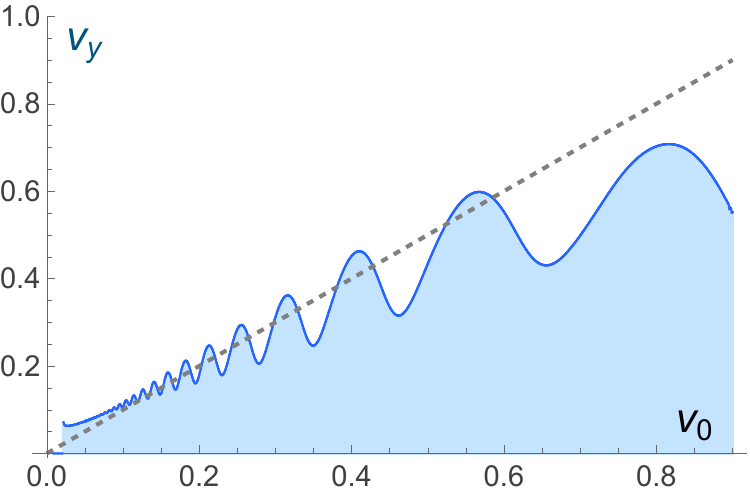} \hspace{0.0cm} \includegraphics[width=4.cm]{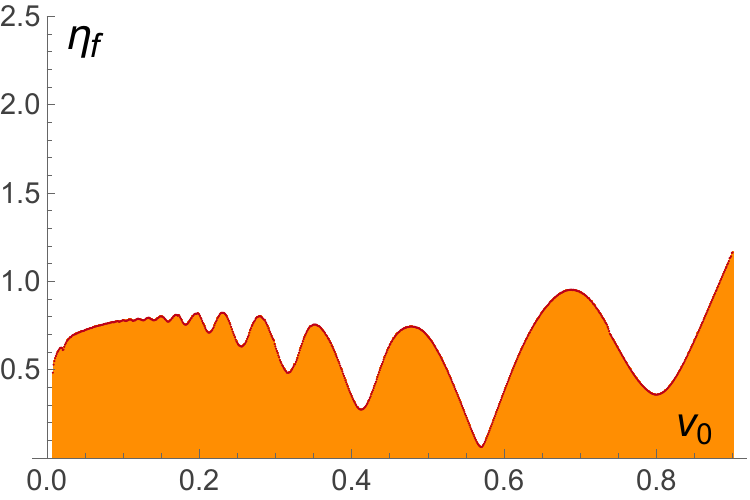} \hspace{0.0cm} \includegraphics[width=4.cm]{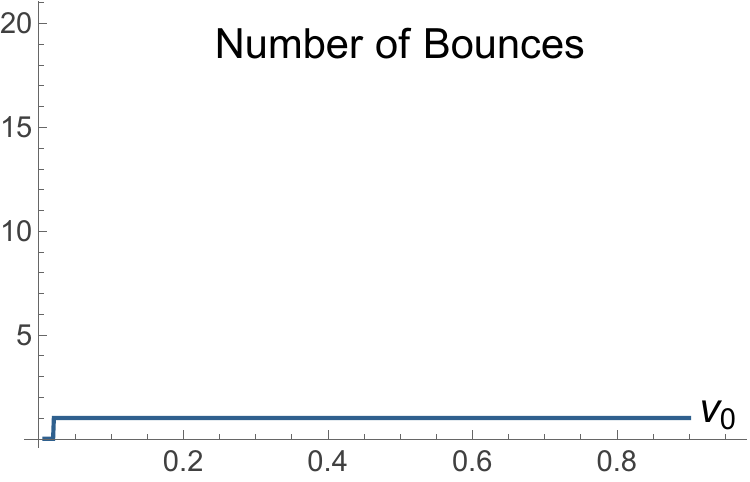}
	\caption{Velocity diagrams for the $v_x$-component (a) and the $v_y$-component (b), together with the amplitude diagram (c) and the number of bounces (d), for head-on scattering of two initially excited 1-vortices with $\lambda=1.2$ and $\eta_0=0.9$.}
	\label{fig:diagramsL12A09}
\end{figure}

We now turn to the case of initially excited vortices. As in the self-dual regime discussed in the previous section, increasing the initial excitation amplitude leads to the emergence of oscillatory patterns in the scattering outputs as functions of the initial velocity. In particular, both the final velocity and the final excitation amplitude exhibit a sequence of local maxima and minima. This behavior is a clear signature of the resonant energy transfer mechanism, whereby energy is periodically exchanged between translational and vibrational degrees of freedom depending on the phase of the oscillation at the instant of collision. This phenomenon is illustrated in Fig. \ref{fig:diagramsL12A09} for the representative case $\eta_0=0.9$.

In addition to the oscillatory structure, two distinctive features arise in the Type-II regime. First, the threshold velocity $v_m$, above which vortices are able to overcome the repulsive interaction and collide, is reduced as the initial excitation amplitude increases. This effect can be understood by considering that two competing forces are present in the system, the intrinsic repulsion characteristic of Type-II vortices and the mode-induced attraction associated with the excitation. Compare Fig. \ref{fig:diagramsL12A00} and \ref{fig:diagramsL12A09}, or see the sequence of graphics in the Appendix as the initial amplitude $\eta_0$ increases.

Secondly, within the low-velocity regime, the final velocity continues to exhibit an oscillatory dependence on the initial one, but it no longer oscillates around the elastic reference line (in contrast to the BPS case). Instead, this oscillatory pattern lies entirely above the line $v_f=v_0$, so that the vortices always characterize superelastic events, where the final velocity exceeds the initial one. This indicates that, in this initial velocity regime, energy transfer is predominantly unidirectional, flowing from the vibrational pool into the kinetic one. Notably, this behavior occurs both below and above the threshold velocity $v_m$, that is, in regimes with and without actual vortex collision, provided that the initial velocity remains sufficiently small. This feature can be clearly observed in Fig. \ref{fig:diagramsL12A09}.

\begin{figure}[h]
	\centering
	\includegraphics[width=4.cm]{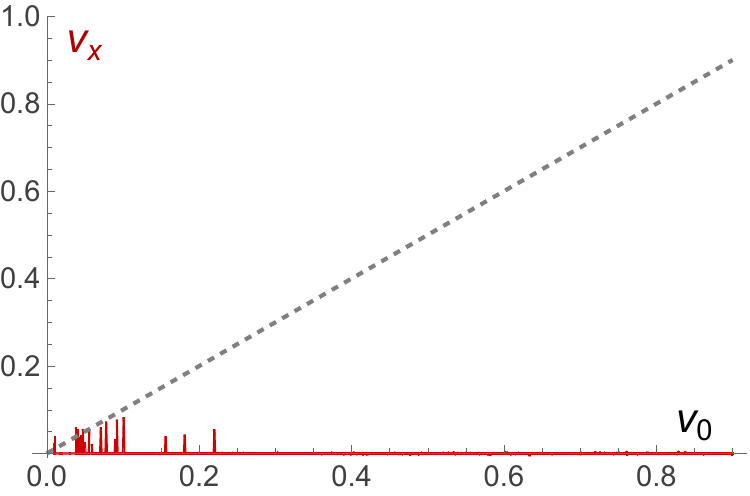} \hspace{0.0cm} \includegraphics[width=4.cm]{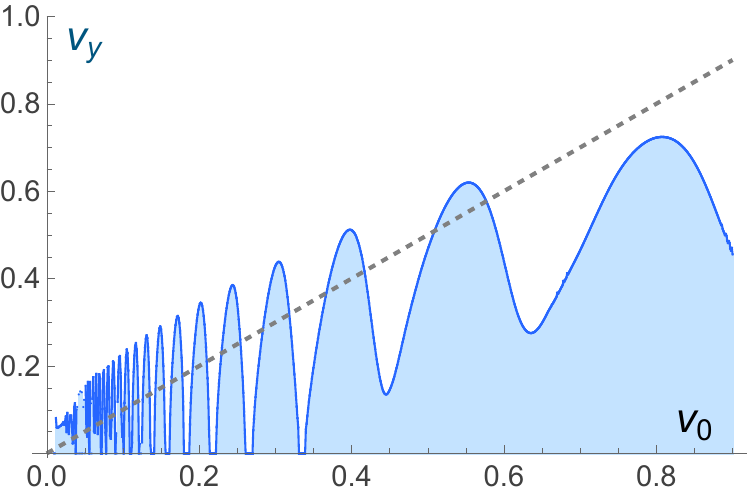} \hspace{0.0cm} \includegraphics[width=4.cm]{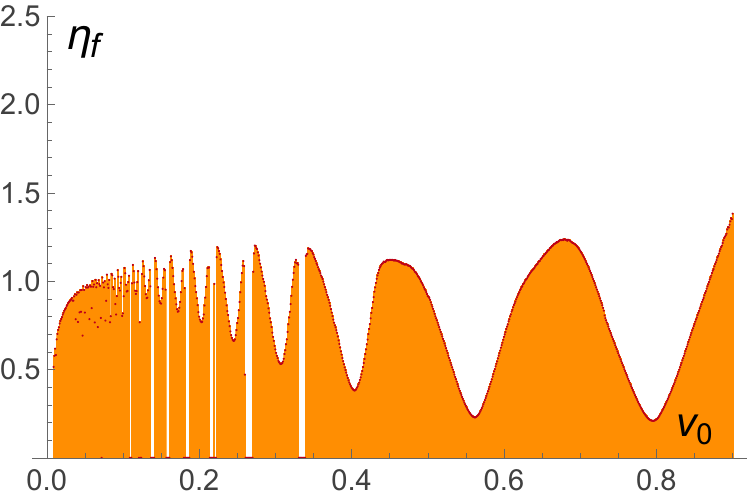} \hspace{0.0cm} \includegraphics[width=4.cm]{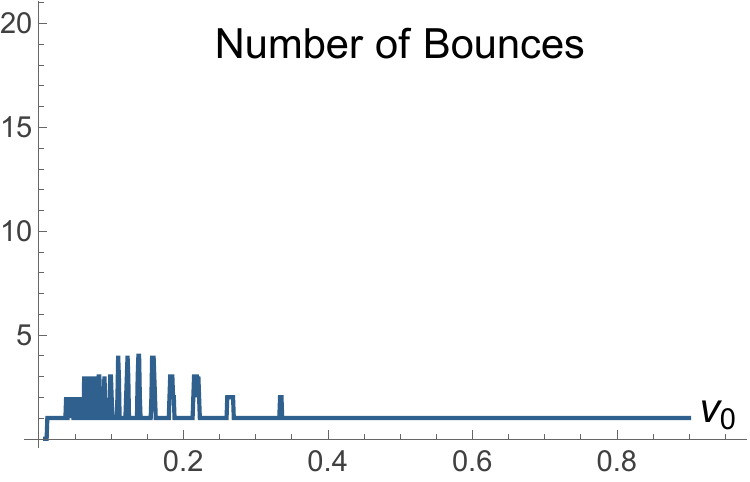}
	\caption{Velocity diagrams for the $v_x$-component (a) and the $v_y$-component (b), together with the amplitude diagram (c) and the number of bounces (d), for head-on scattering of two initially excited 1-vortices with $\lambda=1.2$ and $\eta_0=1.5$.}
	\label{fig:diagramsL12A15}
\end{figure}

Finally, in Fig. \ref{fig:diagramsL12A15} we present an extreme scenario in which the initial excitation amplitude is large, specifically $\eta_0=1.5$. In this regime, the scattering dynamics becomes significantly more intricate. As in the previous subsections, we observe the emergence of isolated 1-bounce windows, separated by intervals where multi-bounce processes take place. Again, the existence of the multi-bounce windows is an effect of the attractive mode-generated force, which temporarily can bind two single vortices. The coexistence of these structures leads to a substantial increase in the complexity of the dynamics. In particular, the scattering diagrams exhibit a proliferation of fractal-like patterns. This behavior highlights the strongly nonlinear and non-adiabatic nature of vortex interactions in the presence of large internal excitations. 

The existence of the attractive mode-generated force and the possibility that, for a sufficiently large amplitude, it overcomes the repulsive static vortex-vortex force can temporarily change the type-II vortices into a system where the vortices effectively behave as type-I vortices.

\section{Conclusions}

The scattering dynamics of wobbling vortices in the Abelian Higgs model is governed by a subtle interplay between static intersoliton forces and dynamical effects arising from internal vibrational modes. The global structure of the scattering diagrams exhibits markedly different phenomenological behaviors for each of the Type-I ($\lambda<1$), BPS ($\lambda=1$) or Type-II ($\lambda>1$) regimes.

In the Type-I regime, vortices experience an intrinsic static attraction, making this case the most dynamically rich and strongly non-linear scenario. Even in the absence of internal excitations, vortex–vortex scattering is characterized by a critical velocity $v_c$, below which the system develops a complex hierarchy of multi-bounce windows, long-lived metastable bound states, and fractal-like structures in the scattering diagram. The inclusion of internal vibrational modes significantly enhances this complexity. As the initial excitation amplitude $\eta_0$ increases, the resonant energy exchange mechanism becomes more efficient, leading to a progressive fragmentation of the one-bounce branch into isolated resonance windows. 

At the self-dual (BPS) point, static forces vanish exactly, providing a clean framework in which to isolate purely mode-driven effects. In the absence of excitations, vortices undergo a universal $90^\circ$-scattering for all initial velocities, with no dependence on a critical threshold. However, when internal modes are excited, the dynamics changes qualitatively. The spectral flow of the vibrational modes induces an effective attractive interaction between vortices. At sufficiently large excitation amplitudes, this induced interaction is strong enough to reproduce structures closely resembling those of the Type-I regime, including multi-bounce events and resonance windows. 

In the Type-II regime, vortices interact via a repulsive force, leading to the emergence of a minimum collision velocity $v_m$ below which vortices cannot overcome the repulsive barrier. In this regime, low-velocity initial conditions typically result in quasi-elastic reflection for unexcited vortices. The presence of internal excitations introduces a competition between repulsive static forces and mode-induced attractive contributions. As a result, the effective threshold velocity $v_m$ decreases with increasing excitation amplitude $\eta_0$, reflecting the ability of vibrational energy to partially compensate for the repulsion and leading to similar multi-bounce patterns as in the previous cases.

\begin{figure}[h]
	\centering
	\includegraphics[width=4.cm]{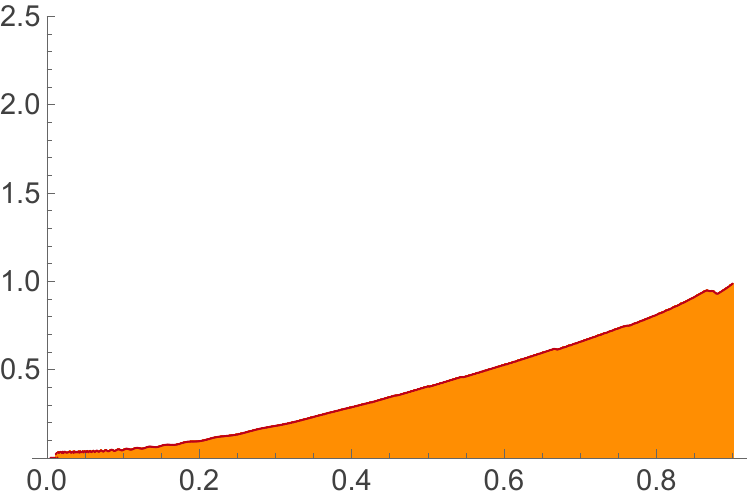} \hspace{0.5cm} \includegraphics[width=4.cm]{SeccionDiagramaAmpliL10A00.pdf} \hspace{0.5cm} \includegraphics[width=4.cm]{SeccionDiagramaAmpliL12A00.pdf} 
	\caption{Amplitude diagrams for head-on scattering of two unexcited 1-vortices with $\lambda=0.8$, $\lambda=1.0$ and $\lambda=1.2$.}
	\label{fig:amplitudes00}
\end{figure}

An important result of our research is that the vortex-vortex collision is a very efficient mechanism for exciting the bound mode of the vortex. Despite having initially two unexcited vortices, after the collision a part of the kinetic energy is transferred to the mode. Furthermore, the strength of this transfer depends on the coupling constant $\lambda$ and the initial velocity $v_0$.  In Fig. \ref{fig:amplitudes00} we present the amplitude of the mode in the final state after the collision of two unexcited vortices in three representative cases $\lambda=0.8, 1, 1.2$. 

The vortices in type-I theory, $\lambda<1$, always excite the bound mode during the scattering, even for an arbitrarily small velocity. This is related to the attractive inter-vortex force that leads to relatively violent processes, independent of the value of $v_0$. Conversely, in the BPS and type-II cases, $\lambda\geq1$, there is a regime of initial velocity in which the collisions are elastic. Thus, the amplitude of the mode is zero in the final state. However, for $v_0$ larger than a critical value (which grows with $\lambda$), the mode is excited and its amplitude increases with the initial velocity. 

There are several physical processes that may be relevant for the excitation of the vortex mode. First, vortices can be excited through interactions with radiation. Second, excited vortices may already be produced during a phase transition. Finally, the mode can also be excited by a thermal bath. It would be interesting to assess the relative importance of these mechanisms. It is worth noting that, in the case of global vortices, i.e., in the limit $\lambda\to \infty$, the last two possibilities are not very efficient \cite{blanco2021}. This may further highlight the role of vortex-vortex collisions.

In any case, the fact that the mode is easily excited in collisions has an obvious consequence for the evolution of a set of many unit-charge vortices. After several two-soliton collisions, the vortices will gradually evolve into their excited versions, transferring more and more kinetic energy to the internal degrees of freedom. Of course, the mode decays into radiation. However, this is a relatively slow process \cite{alonso2024e}. Therefore, one may expect that the evolution of vortices on a medium- or long-time scale will always find them in excited states. In the BPS limit, this means a departure from geodesic dynamics.

The coupling between the kinetic and internal degrees of freedom may have additional consequences for the equation of state of the gas of vortices. It was shown that in the thermodynamic limit a gas of BPS vortices with a finite low density $n=N/V$ has the form of the Clausius equation of a two dimensional non-interacting gas $P(1-4\pi n)=Tn$ \cite{Manton1993}, where the factor $1-4\pi n $ arises from the Bradlow bound \cite{Bradlow1990} and denotes the excluded area of the vortices. Definitely, the energy transfer between the kinetic and internal motion will modify this result. Qualitatively, one can expect the EoS of the van der Waals form, similar to what happens after the inclusion of quantum corrections \cite{Manton2022}. Undoubtedly, devoted studies are required to solve this problem. 

Another direction in which our work can be continued is the case of strongly type-II vortices with $\lambda \ge 1.5$. Although there is no genuine bound mode, there are some excitations that may participate in the resonant energy transfer mechanism. These are the half-threshold mode $(\lambda=1.5)$ and half-bound modes, i.e., Feshbach resonances $(\lambda >1.5)$. Both may lead to the appearance of nonlinear oscillon-like excitation on the vortex \cite{Blaschke2026}. Furthermore, the lowest Feshbach resonance is known to play a distinguished role in vortex-antivortex annihilation \cite{bachmaier2026}. Hence, we expect that the two-vortex collision also in this regime may reveal a nontrivial structure.

\appendix

\section{Parameter scan of scattering diagrams across the $\lambda - \eta_0$ plane}

This appendix introduces a systematic graphical overview of the scattering dynamics across the parameter space explored in this work. The first table collects the velocity scattering diagrams, where the dependence of the final vortex velocities on the initial collision velocity is displayed for different values of the coupling constant $\lambda$ and the initial excitation amplitude $\eta_0$. The second table contains the corresponding amplitude scattering diagrams, in which the final internal excitation of the vortices is shown as a function of the initial velocity. These figures make explicit the role of energy exchange between translational and internal degrees of freedom, and provide a direct way to track how vibrational energy is redistributed during the collision process across the different regimes.

\begin{figure}[h]
	\begin{tabular}{c|ccc}
		& $\lambda=0.5$ & $\lambda=1.0$ & $\lambda=1.2$  \\ \hline
		\rotatebox{90}{\hspace{0.8cm} $\eta_0= 0.0$} &\includegraphics[width=4.5cm]{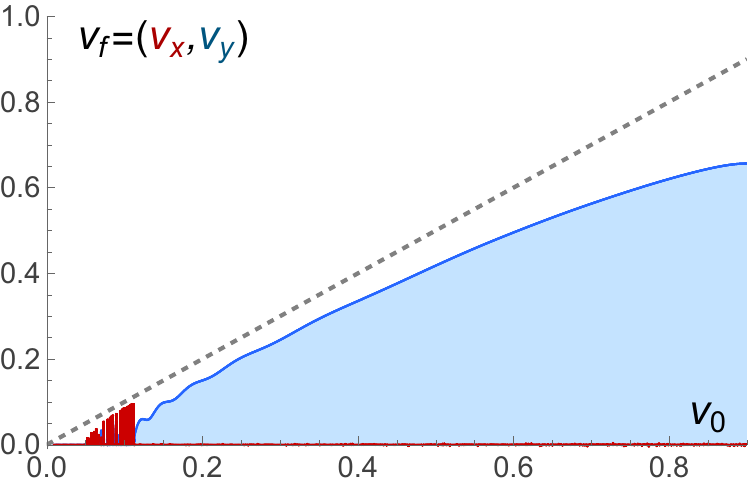} & \includegraphics[width=4.5cm]{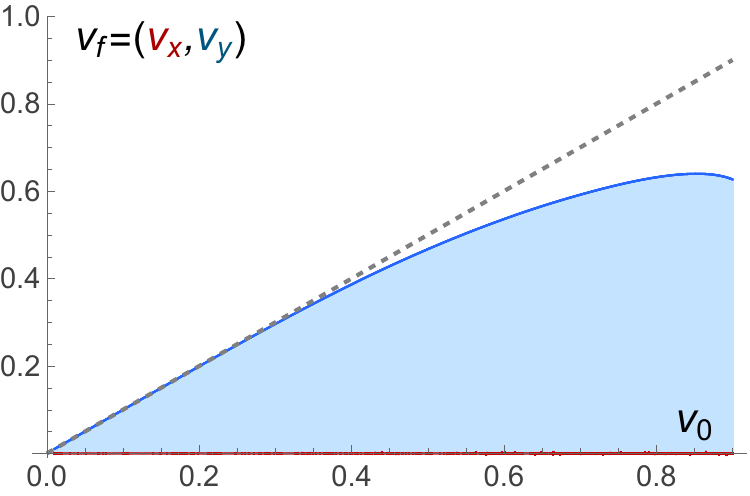} & \includegraphics[width=4.5cm]{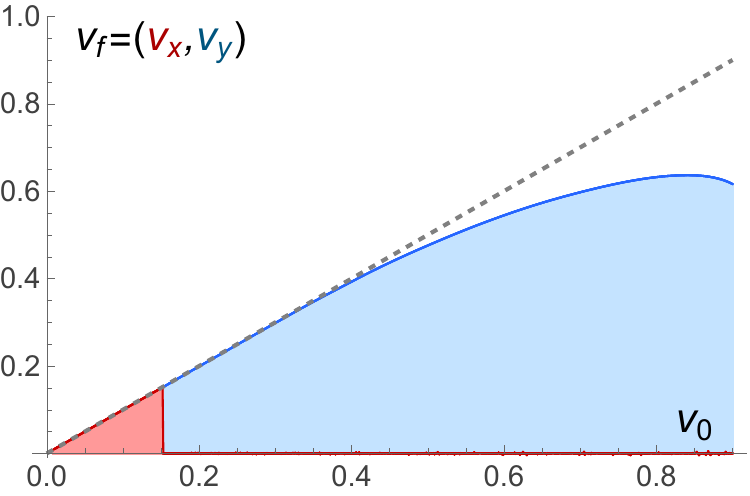}  \\
		\rotatebox{90}{\hspace{0.8cm} $\eta_0= 0.3$} &\includegraphics[width=4.5cm]{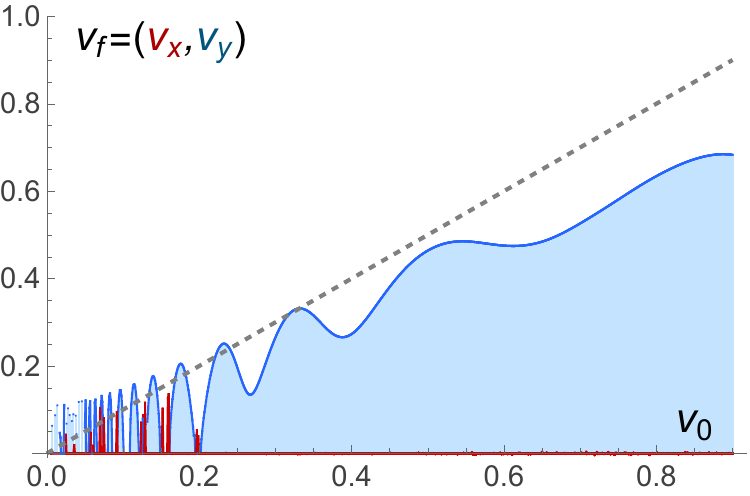} & \includegraphics[width=4.5cm]{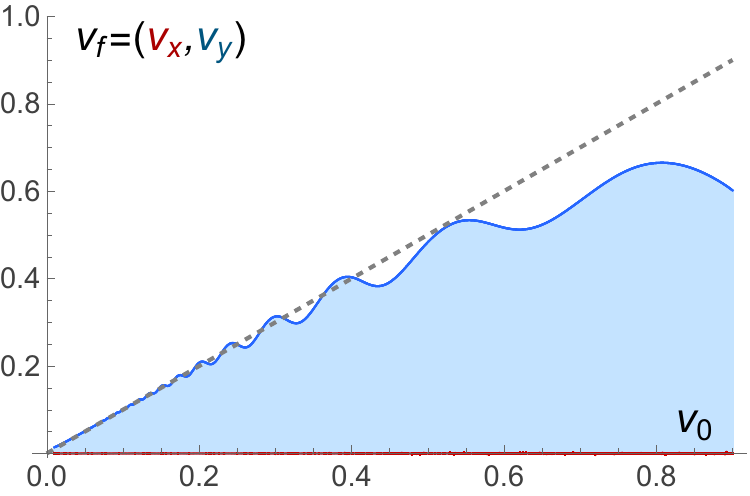} & \includegraphics[width=4.5cm]{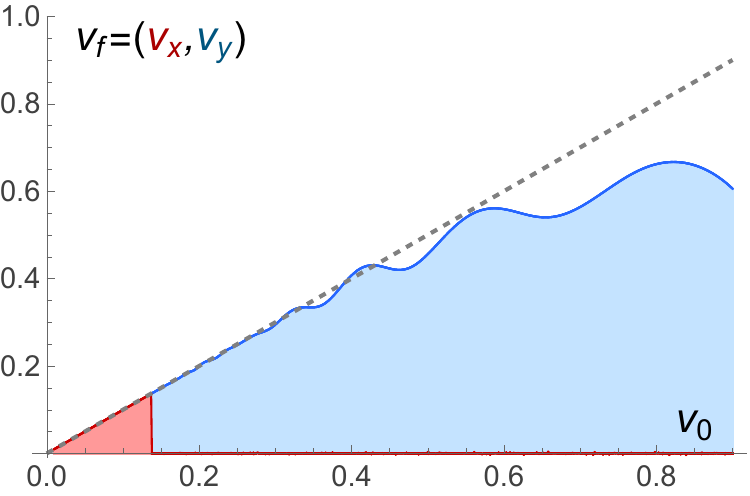} \\
		\rotatebox{90}{\hspace{0.8cm} $\eta_0= 0.6$} &\includegraphics[width=4.5cm]{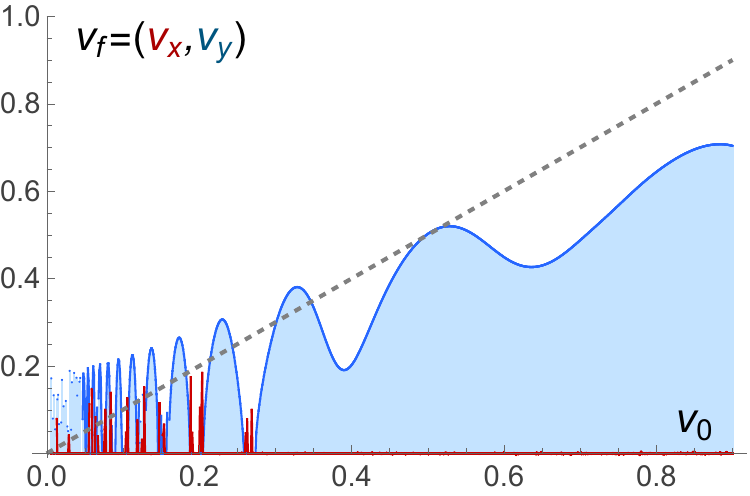} & \includegraphics[width=4.5cm]{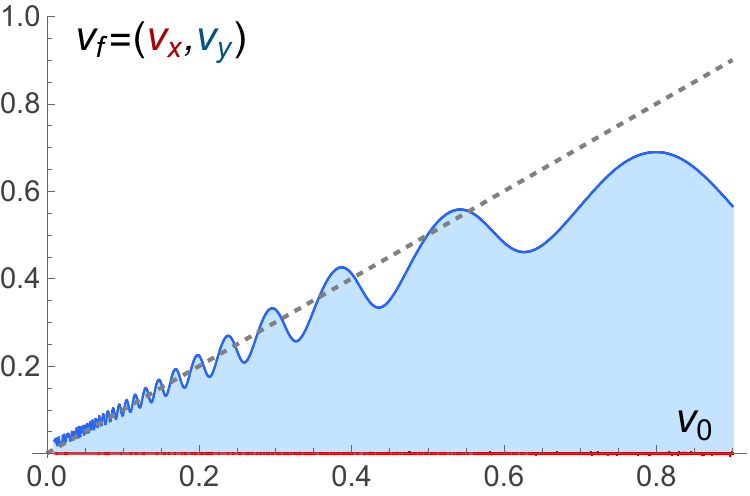} & \includegraphics[width=4.5cm]{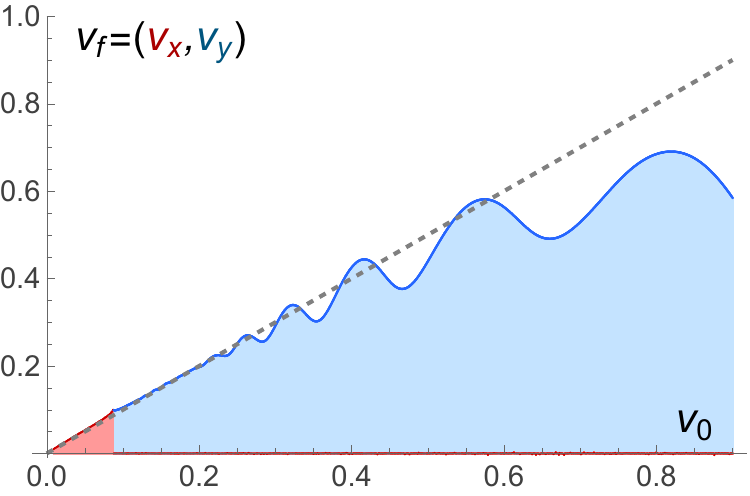} \\
		\rotatebox{90}{\hspace{0.8cm} $\eta_0= 0.9$} &\includegraphics[width=4.5cm]{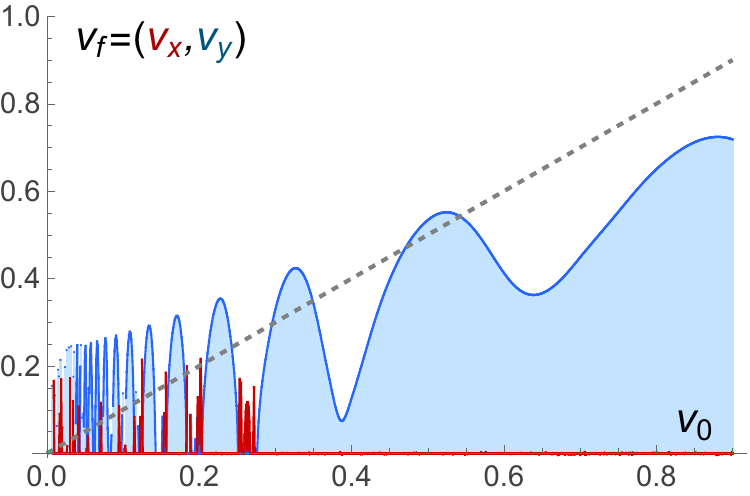} & \includegraphics[width=4.5cm]{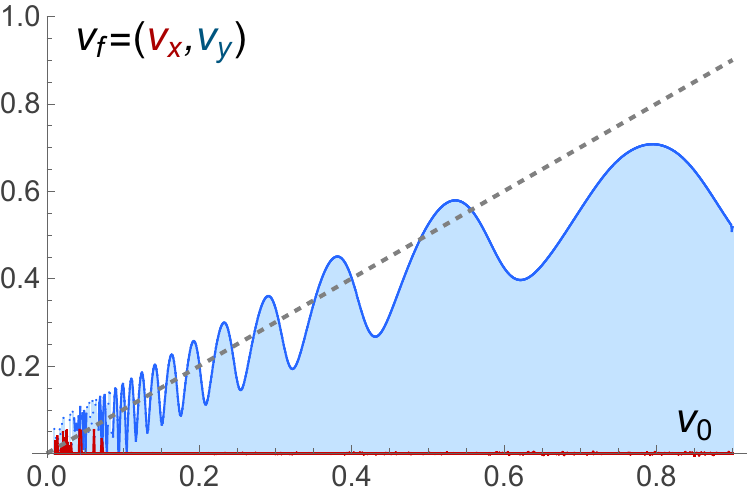} & \includegraphics[width=4.5cm]{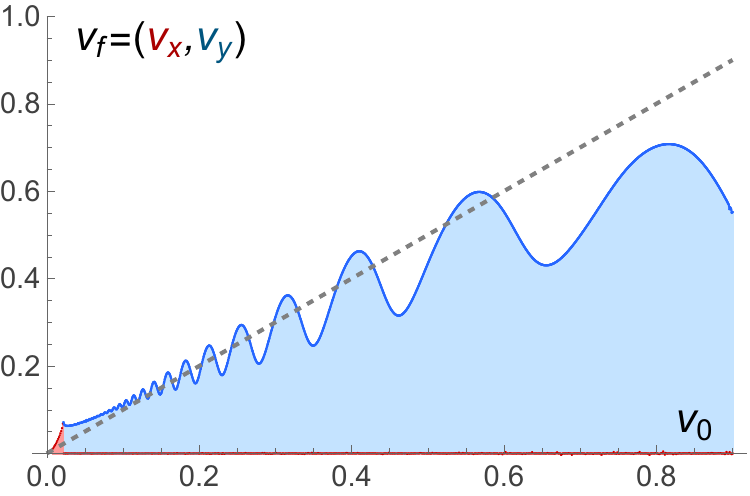} \\
		\rotatebox{90}{\hspace{0.8cm} $\eta_0= 1.2$} &\includegraphics[width=4.5cm]{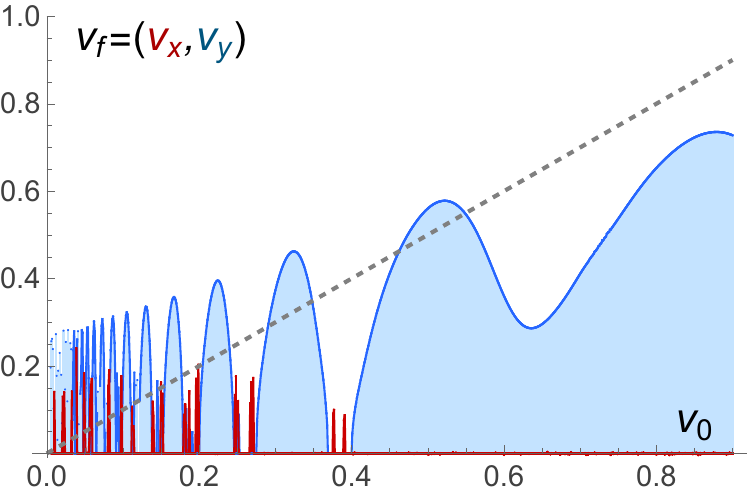} & \includegraphics[width=4.5cm]{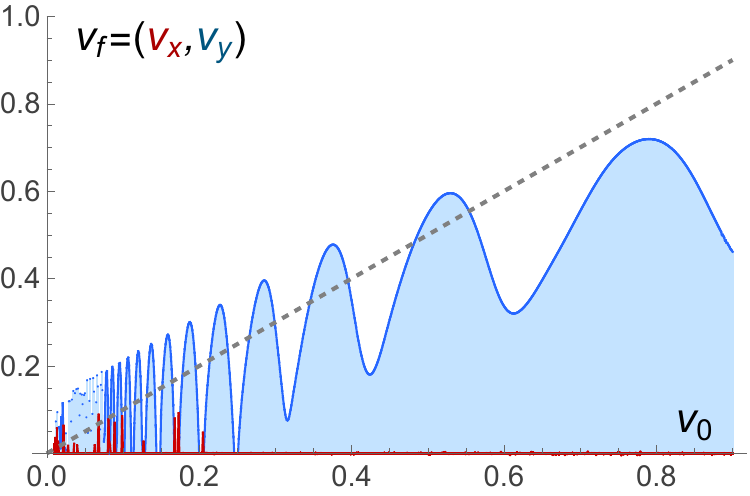} & \includegraphics[width=4.5cm]{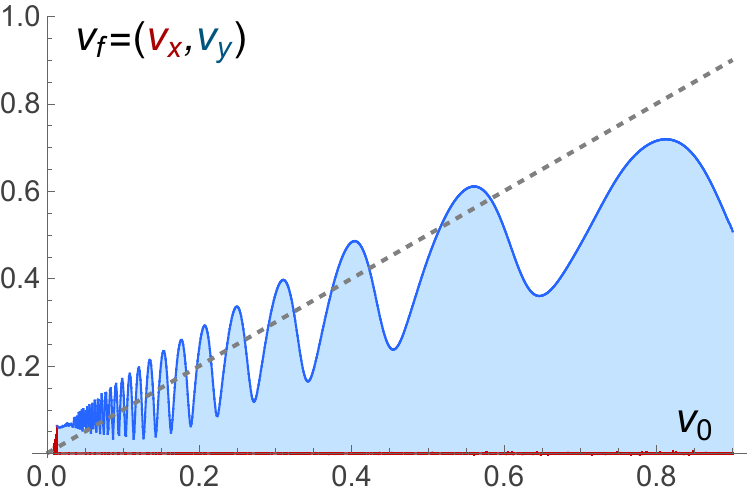} \\
		\rotatebox{90}{\hspace{0.8cm} $\eta_0= 1.5$} &\includegraphics[width=4.5cm]{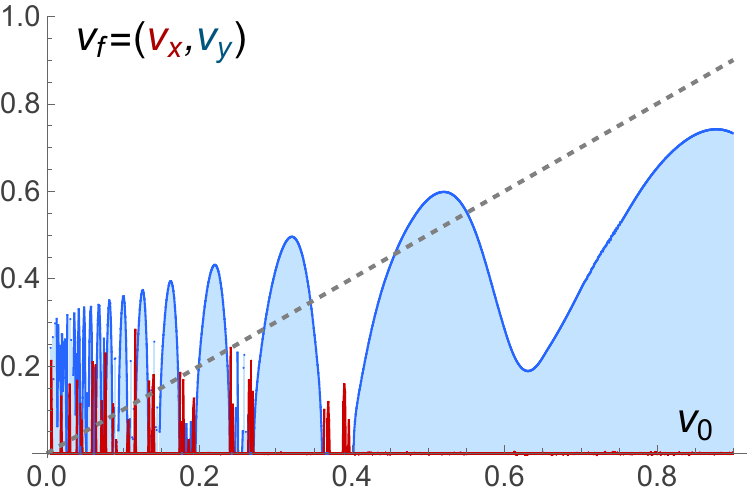} & \includegraphics[width=4.5cm]{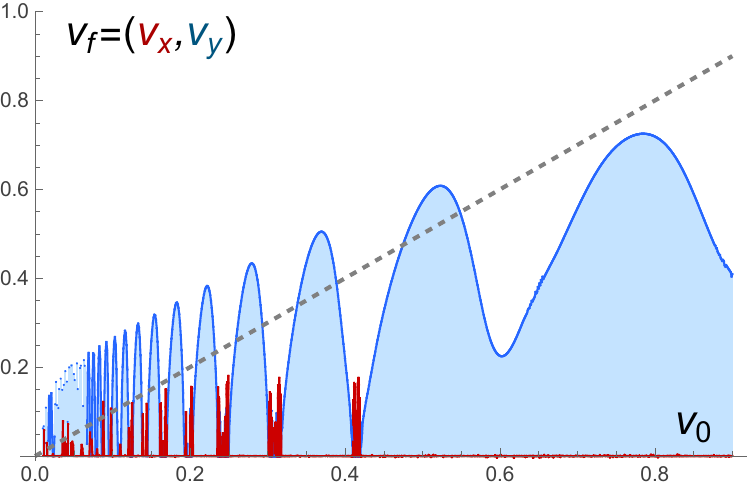} & \includegraphics[width=4.5cm]{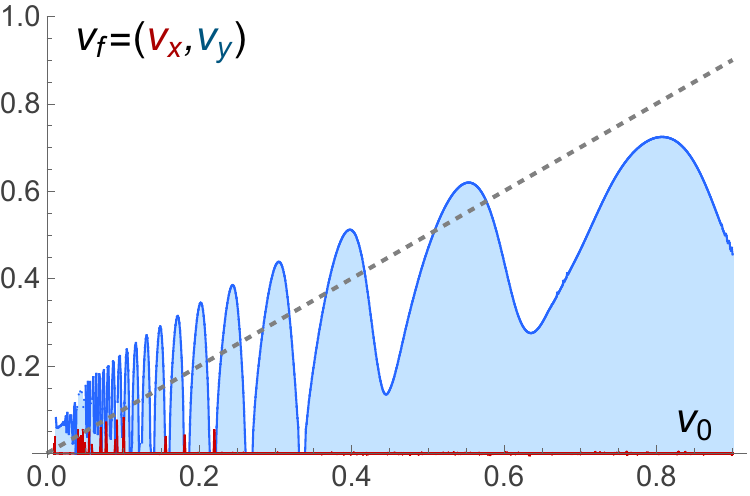} \\
	\end{tabular}
	\caption{Velocity scattering diagrams for head-on collisions of two excited 1-vortices. Columns correspond to $\lambda=0.5$, $1.0$, and $1.2$, while rows correspond to initial excitation amplitudes $\eta_0=0.0$, $0.3$, $0.6$, $0.9$, $1.2$, and $1.5$.}
\end{figure}

\begin{figure}[h]
	\begin{tabular}{c|ccc}
		& $\lambda=0.5$ & $\lambda=1.0$ & $\lambda=1.2$  \\ \hline
		\rotatebox{90}{\hspace{0.8cm} $\eta_0= 0.0$} &\includegraphics[width=4.5cm]{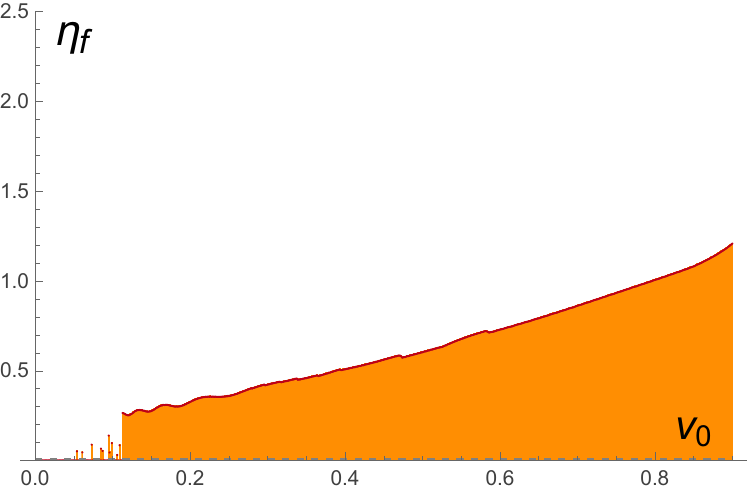} & \includegraphics[width=4.5cm]{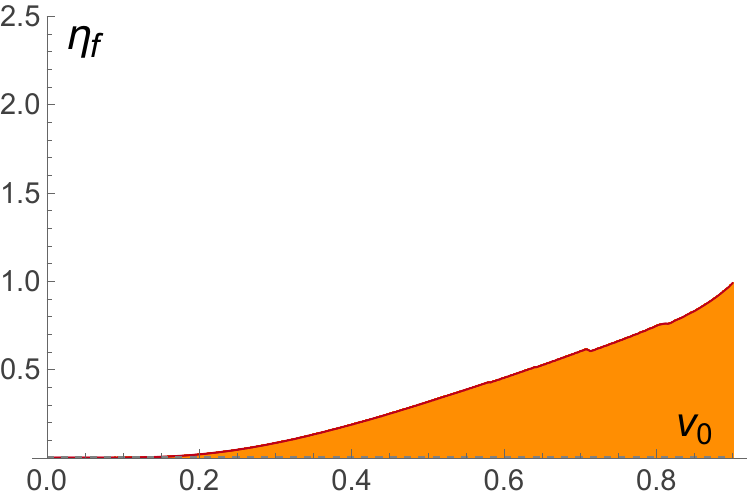} & \includegraphics[width=4.5cm]{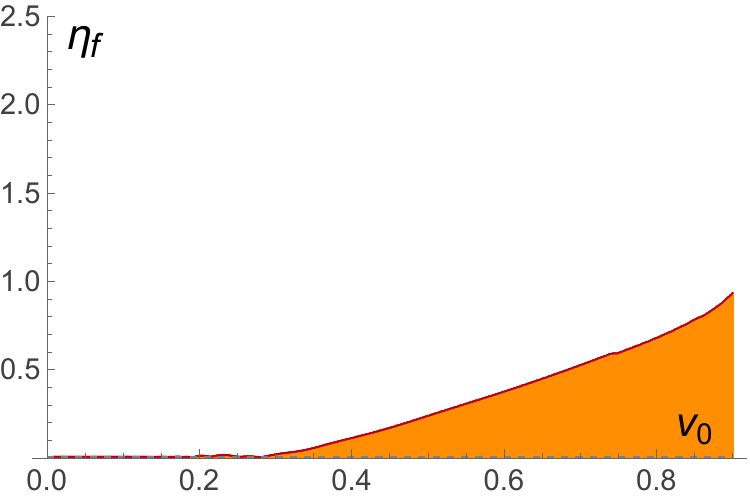}  \\
		\rotatebox{90}{\hspace{0.8cm} $\eta_0= 0.3$} &\includegraphics[width=4.5cm]{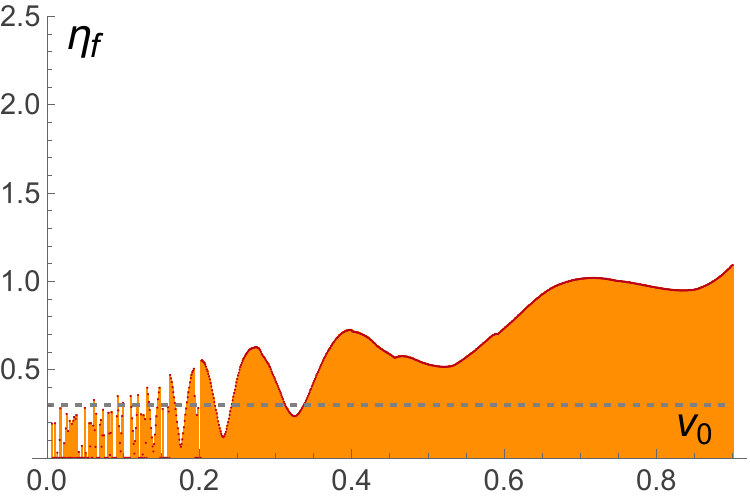} & \includegraphics[width=4.5cm]{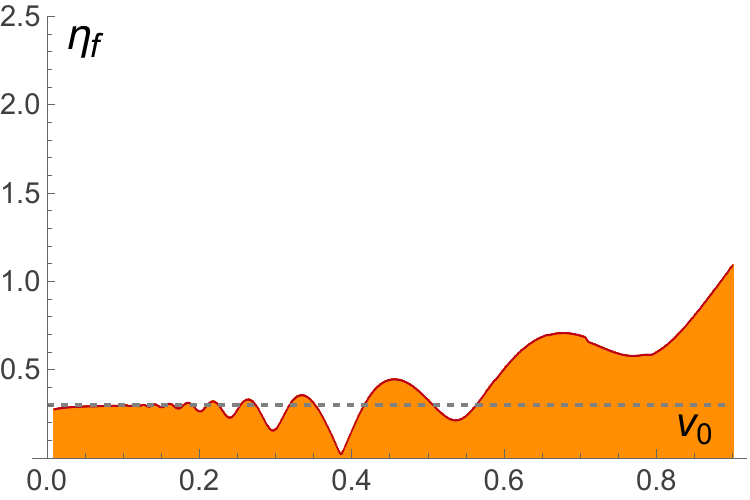} & \includegraphics[width=4.5cm]{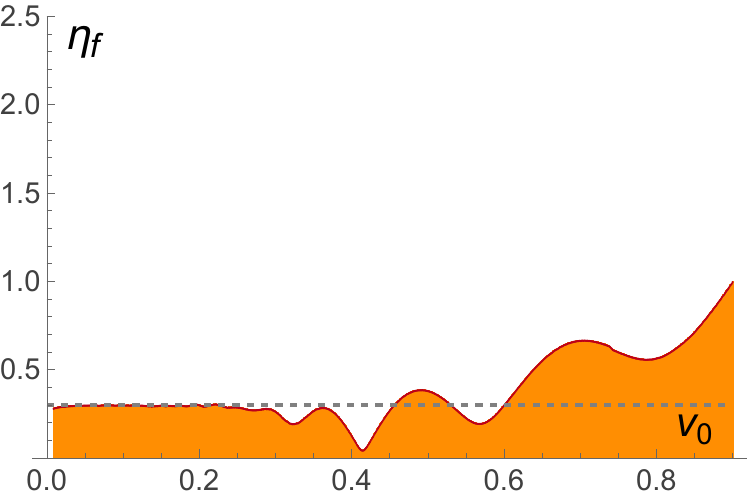} \\
		\rotatebox{90}{\hspace{0.8cm} $\eta_0= 0.6$} &\includegraphics[width=4.5cm]{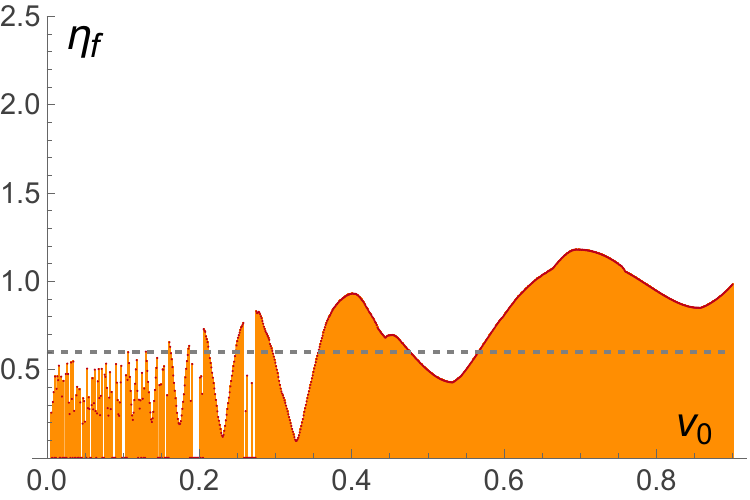} & \includegraphics[width=4.5cm]{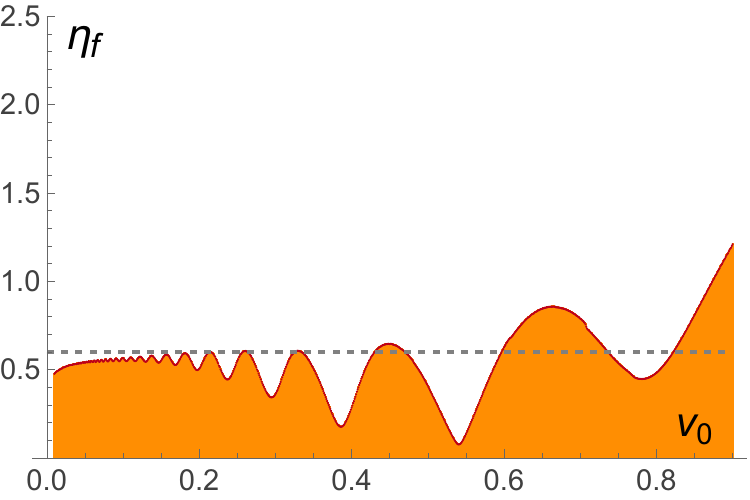} & \includegraphics[width=4.5cm]{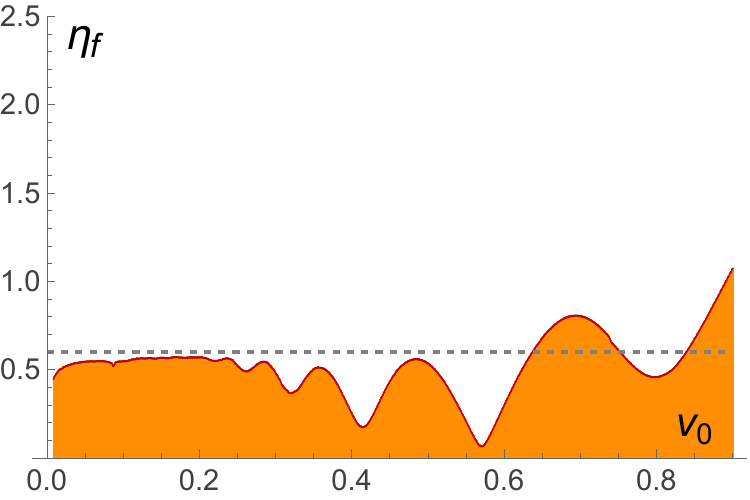} \\
		\rotatebox{90}{\hspace{0.8cm} $\eta_0= 0.9$} &\includegraphics[width=4.5cm]{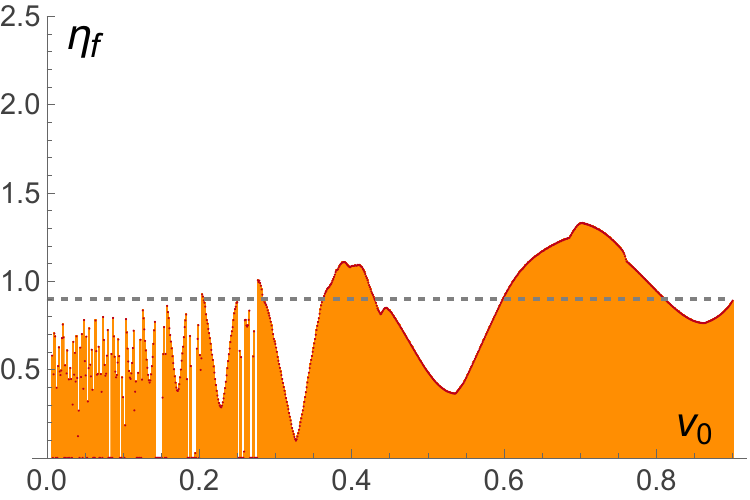} & \includegraphics[width=4.5cm]{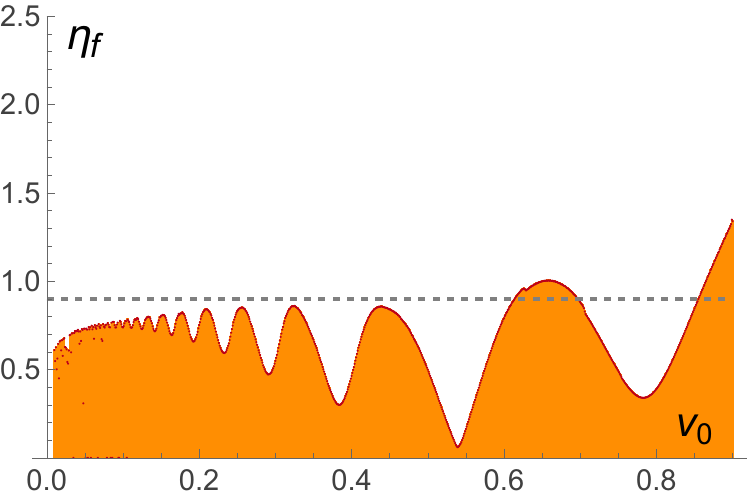} & \includegraphics[width=4.5cm]{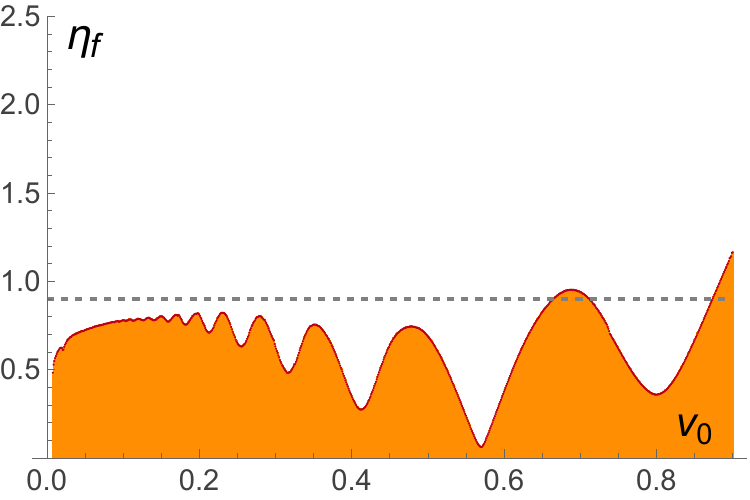} \\
		\rotatebox{90}{\hspace{0.8cm} $\eta_0= 1.2$} &\includegraphics[width=4.5cm]{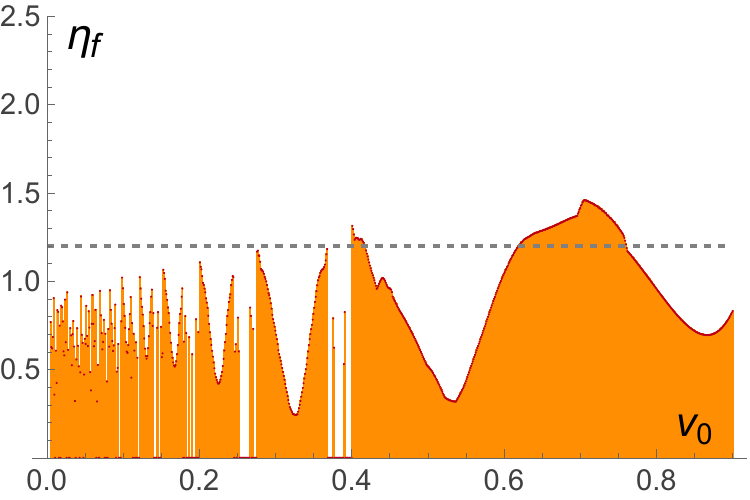} & \includegraphics[width=4.5cm]{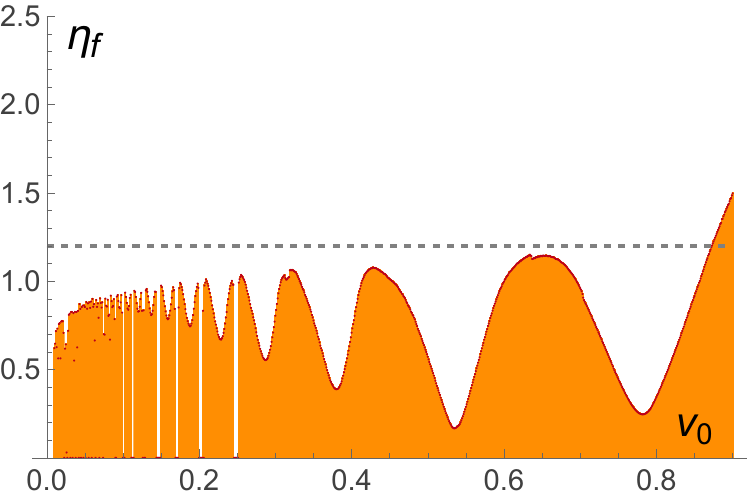} & \includegraphics[width=4.5cm]{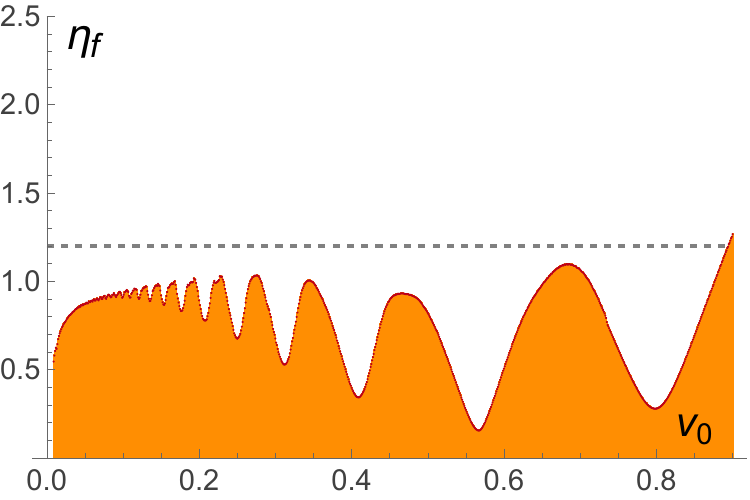} \\
		\rotatebox{90}{\hspace{0.8cm} $\eta_0= 1.5$} &\includegraphics[width=4.5cm]{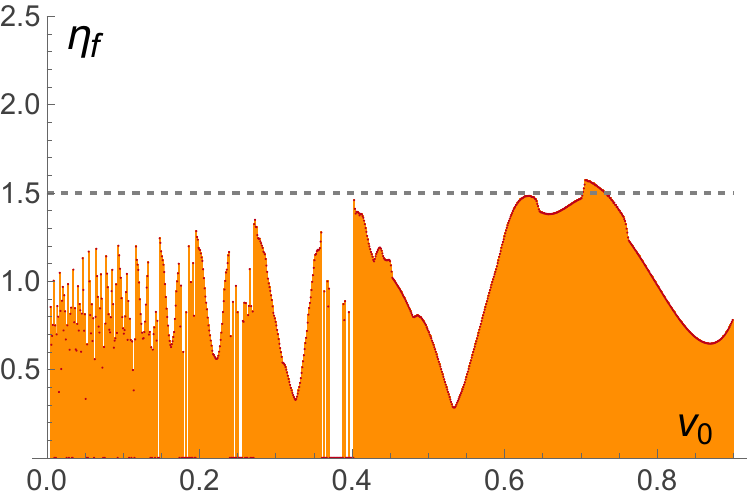} & \includegraphics[width=4.5cm]{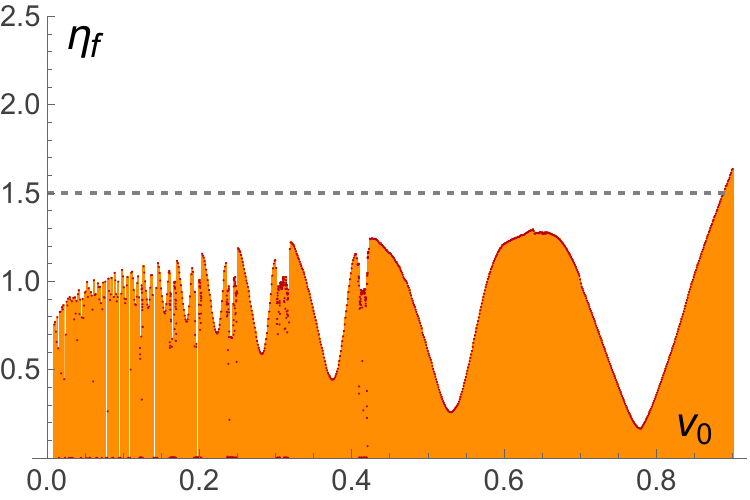} & \includegraphics[width=4.5cm]{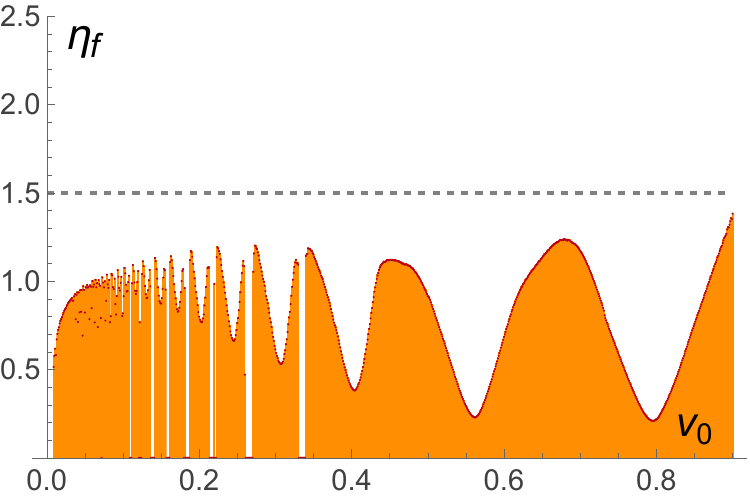} \\
	\end{tabular}
	\caption{Amplitude scattering diagrams for head-on collisions of two excited 1-vortices. Columns correspond to $\lambda=0.5$, $1.0$, and $1.2$, while rows correspond to initial excitation amplitudes $\eta_0=0.0$, $0.3$, $0.6$, $0.9$, $1.2$, and $1.5$.}
\end{figure}

%%%%%%%%%%%%%%%%%%%%%%%%%%%%%%%%%%%%%
\section*{Acknowledgments}
%%%%%%%%%%%%%%%%%%%%%%%%%%%%%%%%%%%%%

The authors have been supported in part by Spanish Ministerio de Ciencia e Innovación (MCIN) with funding from the grant PID2023-148409NB-I00 MTM, and they were also partially supported by Fundaci\'on Sol\'orzano through the project FS/11-2024.

\section*{Conflict of Interest}
The authors declare that they have no conflict of interest.

\section*{Data availibility}
Data sharing is not applicable to this article as no datasets were generated or analysed during
the current study.

\end{document}